\documentclass[%
 aip,
rsi,%
 amsmath,amssymb,
 reprint,%
]{revtex4-1}
\pdfoutput=1 
\usepackage{graphicx}
\usepackage{dcolumn}
\usepackage{bm}
\usepackage{graphicx}
\usepackage{bm}
\usepackage{amsmath} 
\usepackage{float} 
\usepackage{dcolumn}
\usepackage[utf8x]{inputenc}
\usepackage[T1]{fontenc}
\usepackage[colorlinks=true, allcolors=blue]{hyperref}
\usepackage[normalem]{ulem}
\useunder{\uline}{\ul}{}

\begin{document}


\title[Behavior and Breakdown of Higher-Order Fermi-Pasta-Ulam-Tsingou Recurrences]{Behavior and Breakdown of Higher-Order Fermi-Pasta-Ulam-Tsingou Recurrences}

\author{Salvatore D. Pace and David K. Campbell}
\affiliation{Department of Physics, Boston University, Boston, Massachusetts 02215, USA}

\date{\today}

\begin{abstract}
We investigate numerically the existence and stability of higher-order recurrences (HoRs), including super-recurrences, super-super-recurrences, \textit{etc.}, in the $\alpha$ and $\beta$ Fermi-Pasta-Ulam-Tsingou (FPUT) lattices for initial conditions in the fundamental normal mode. Our results represent a considerable extension of the pioneering work of Tuck and Menzel on super-recurrences. For fixed lattice sizes, we observe and study apparent singularities in the periods of these HoRs, speculated to be caused by nonlinear resonances. Interestingly, these singularities depend very sensitively on the initial energy and the respective nonlinear parameters. Furthermore, we compare the mechanisms by which the super-recurrences in the two models breakdown as the initial energy and respective nonlinear parameters are increased. The breakdown of super-recurrences in the $\beta$-FPUT lattice is associated with the destruction of the so-called metastable state and hence is associated with relaxation towards equilibrium. For the $\alpha$-FPUT lattice, we find this is not the case and show that the super-recurrences break down while the lattice is still \textit{metastable}. We close with comments on the generality of our results for different lattice sizes. 
\end{abstract}

\maketitle

\begin{quotation}
Since 1953, Fermi-Pasta-Ulam-Tsingou (FPUT) recurrences have raised many questions regarding how FPUT lattices actually approach equilibrium. The subtleties involved in these recurrences have been studied for decades, and from these rich projects arose new fields of physics and mathematics. Alongside FPUT recurrences are the super-recurrences first studied by Tuck and Menzel. They amount to a periodic modulation of the FPUT recurrences, in which an even greater amount of energy is returned to the initial state. These complex behaviors are defining features for the out of equilibrium behavior exhibited in FPUT lattices. Our study consists of a considerable extension of the pioneering work done by Tuck and Menzel. We have studied recurrences which include the original FPUT recurrences and super-recurrences, but also \textit{higher-order recurrences} (HoRs) like super-super-recurrences. We have investigated the nontrivial behavior and breakdown of HoRs in both the $\alpha$ and $\beta$ FPUT lattices, studying their periods as a function of energy, and the relationship between HoRs and the so-called \textit{metastable} state. The interesting differences found between the $\alpha$ and $\beta$-FPUT lattices provide further evidence of the subtleties involved in the approach to equilibrium at low energies in classical many-body systems.    
\end{quotation}

\section{Introduction}\label{Introduction}
\begin{figure}[t!]
	\includegraphics[width=.48\textwidth]{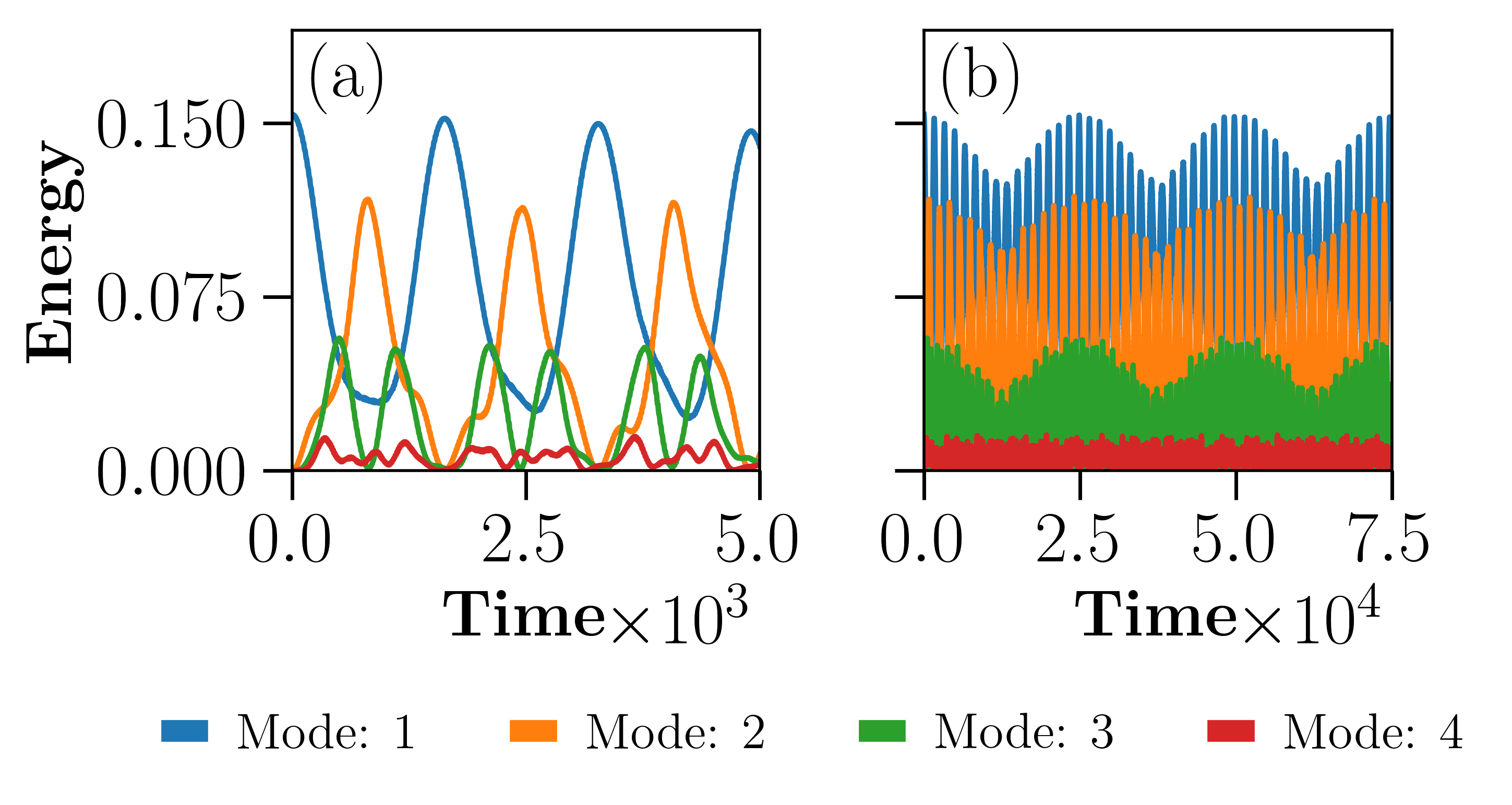}
	\caption{(a) Recurrences ($\langle T_{R} \rangle\sim 1.66\times10^{3}$) and (b) super-recurrences ($\langle T_{SR} \rangle\sim 2.53\times10^{4}$) are shown for the $\alpha$-FPUT lattice ($N=15$, $E=0.15372$, and $\alpha = 0.25$).} 
	\label{arecurrences}
	\bigbreak
	\includegraphics[width=.48\textwidth]{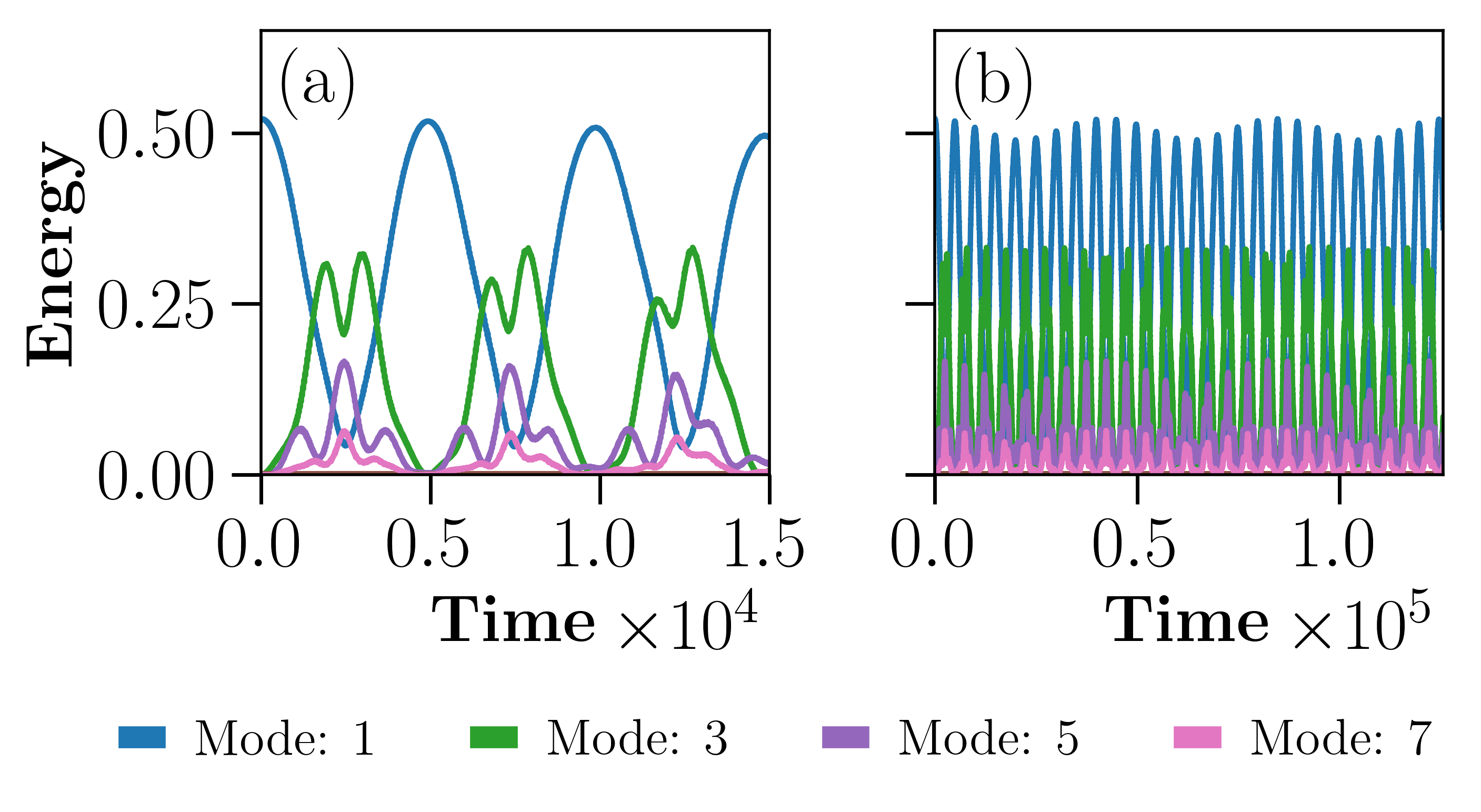}
	\caption{(a) Recurrences ($\langle T_{R} \rangle \sim 4.99\times10^{3}$) and (b) super-recurrences ($\langle T_{SR} \rangle \sim 4.19\times10^{4}$) are shown for the $\beta$-FPUT lattice ($N=31$, $E=0.5208$, and $\beta=1$).} 
	\label{brecurrences}
\end{figure}

In 1953, Fermi, Pasta, Ulam, and Tsingou \cite{FPUT} (FPUT) began computer-based numerical simulations to investigate the rate of thermalization in a nonlinear lattice of equi-mass, anharmonic oscillators with initial conditions far from equilibrium. They expected to see the system quickly reach equipartition due to modal couplings caused by nonlinearities. However, for their initial conditions, they famously observed that energy was shared among only a few of the lowest normal modes and that remarkable near-recurrences of the initial state existed, seemingly contradicting the expectation of equipartition \cite{fpu}. These FPUT \textit{recurrences} and the questions they raised about how equilibrium is actually approached have been the focus of numerous studies \cite{zabusky,ford,dauxois, gallavotti,50yrs,exp maths, Weissert}, have given rise to "soliton" theory \cite{zabusky and kruskal, zabusky and deem}, and continue to challenge researchers today \cite{lin,ermoshin, flach,nelson,bruggeman}. Our recalculations of some examples of the original FPUT recurrences are shown in Figs \ref{arecurrences}a and \ref{brecurrences}a.

One of the obvious first objections to the FPUT study was that they had not run the system long enough to allow it to equilibrate. This objection was investigated by Tuck and Menzel (née Tsingou), beginning in the 1960's with a series of much longer computational runs, the results of which were published in 1972 \cite{TandM}. Tuck and Menzel discovered that for longer-time computer runs, instead of equipartition, the initial conditions chosen by FPUT produced \textit{super-recurrences} (SRs) in which a still greater fraction of the initial energy returned to the initial state. In essence, their results showed that these SRs amounted to a periodic modulation of the original FPUT recurrence as shown in Figs. \ref{arecurrences}b and \ref{brecurrences}b. 

A first possible explanation for these SRs was given in 1965 by Zabusky and Kruskal who learned of SRs from private communications with Tuck. When they first observed solitons causing recurrences in the Korteweg–de Vries (KdV) equation, which they considered as a continuum limit for the $\alpha$-FPUT lattice, they conjectured that solitons interacting more and more and then less and less out of phase at the recurrence times causes SRs \cite{zabusky and kruskal}. 
However, for many years after Tuck and Menzel published their results, SRs in FPUT-lattices were not further pursued. Although there were citations that referred to the history of Tuck and Menzel's work in the 1970s and early 1980s, their simulations were not revisited in detail until the 1987 paper of Drago and Ridella \cite{drago and ridella}.  Drago and Ridella were able to replicate the SRs for the $\alpha$-FPUT lattice but claimed that Tuck and Menzel's SRs for the $\beta$-FPUT lattice were a numerical artifact due to their time step size having been too large. In 1991, Sholl and Henry \cite{henry and sholl} approached both the FPUT recurrences and SRs analytically through a shifted-frequency perturbation scheme. They concluded that the SRs in the $\alpha$-FPUT lattice are due to a beat-like mechanism from different resonances among nonlinear frequencies. However, they could not find a general explanation for the SR in the $\beta$-FPUT lattice. Nevertheless, they proposed that the SRs in the $\beta$-FPUT lattice, and in general \textit{other higher-order recurrences} (HoRs) (i.e. super-super-recurrences which are a periodic modulation of the super-recurrences) could exist but may be hard to detect numerically. Another interpretation of SRs has been given by Weissert, who while discussing the Kolmogorov-Arnold-Moser (KAM) theorem \cite{Weissert}, gave a heuristic argument for the existence of SRs in terms of KAM tori in phase space.

FPUT-like recurrences and SRs have also been observed and studied in other models. FPUT recurrences has been studied in the nonlinear Schrödinger equation \cite{NLSE_R,NLSE_R2}, interacting Bose gases \cite{bose}, and in the dynamics of anti-de Sitter space \cite{AdS}. FPUT recurrences have also been observed experimentally in deep water-waves \cite{water_FPUT}, optical fibers \cite{optical_FPUT}, and a feedback ring system \cite{ring_FPUT}. Further studies in KdV type equations have observed and studied SRs in terms of what is called the dispersion parameter \cite{solition_SR,solition_SR_2,solition_SR_3}. Recently, SRs have been observed in the nonlinear Schrödinger equation for initial states made up of multiple unstable modes \cite{NLSE_2}.

In this paper, we extend previous studies by numerically investigating HoRs in both the $\alpha$ and $\beta$-FPUT lattices with zero momenta, fundamental mode initial states. To clarify our terminology, HoRs include any recurrences that exist other than the usual FPUT recurrences. To generalize the idea of recurrence, we will be referring to the $n^{\text{th}}$-order recurrences. First-order recurrences (1oRs) will be the normal FPUT recurrences, second-order recurrences (2oRs) will refer to Tuck and Menzel's super-recurrences, third-order recurrences (3oRs) will refer to super-super-recurrences, and so forth. In terms of HoRs, $n^{\text{th}}$-order recurrences ($n$oRs) amount to periodic modulations of $(n-1)^{\text{th}}$-order recurrences ($(n-1)$oRs). 

In section \ref{The Model}, we introduce the $\alpha$ and $\beta$-FPUT lattices along with a rescaling of normal mode coordinates and momenta to show that the numerical results to depend only on the quantities $E\alpha^{2}$ and $E\beta$, in the respective models \cite{De Luca}. Then, in section \ref{revist}, we revisit the previous results on 2oRs in FPUT lattices \cite{TandM,drago and ridella,henry and sholl} to set the background and context for our study. Section \ref{HoR exists} shows the existence of HoRs, greater than second order, in both the $\alpha$ and $\beta$-FPUT lattices. In section \ref{HoR}, we study apparent singularities in the HoRs periods and show that a nested set of HoRs exists at these apparent divergences. Then in section \ref{breakdown}, we explore how the 2oRs in the $\alpha$ and $\beta$-FPUT lattices breakdown and relate these breakdown mechanisms to the previously observed \textit{metastable states} \cite{Benettin}. Finally, in section \ref{N}, we explore the dependence of our results on the size ($N$) of the FPUT lattice and show that our results are essentially independent of $N$ for large $N$. 

\section{The Fermi-Pasta-Ulam-Tsingou Lattices}\label{The Model}
Treating $N$ to be the number of \textit{active} anharmonic oscillators \cite{note on N}, the Hamiltonian of the $\alpha$-FPUT lattice is given by
\begin{equation} \label{afpuham}
H_{\alpha}(\bm{q},\bm{p}) = \sum_{n=1}^{N}\frac{p_{n}^{2}}{2} + \sum_{n=0}^{N}\frac{1}{2}(q_{n+1}-q_{n})^{2}+\frac{\alpha}{3}(q_{n+1}-q_{n})^{3},
\end{equation}
while the Hamiltonian of the $\beta$-FPUT lattice is given by
\begin{equation} \label{bfpuham}
H_{\beta}(\bm{q},\bm{p}) = \sum_{n=1}^{N}\frac{p_{n}^{2}}{2} + \sum_{n=0}^{N}\frac{1}{2}(q_{n+1}-q_{n})^{2}+\frac{\beta}{4}(q_{n+1}-q_{n})^{4},
\end{equation}
both with fixed boundary conditions $q_{0}=q_{N+1}=0$ and $p_{0}=p_{N+1}=0$, where $q_{n}(t)$ and $p_{n}(t)$ are canonical coordinates and momenta, respectively. The normal modes can be investigated through the involutorial canonical transformation
\begin{equation}
\begin{bmatrix}
q_{n} \\
p_{n} \\
\end{bmatrix}
= \sqrt{\frac{2}{N+1}}\sum_{k=1}^{N} 
\begin{bmatrix}
Q_{k} \\
P_{k} \\
\end{bmatrix}
\sin\left({\frac{nk\pi}{N+1}}\right),
\end{equation}
which diagonalizes the harmonic oscillator Hamiltonian. Rewriting equations \ref{afpuham} and \ref{bfpuham} in these normal mode coordinates ($\bm{Q},\bm{P}$) yields 
\begin{equation}\label{modalaham}
H_{\alpha}(\bm{Q},\bm{P}) = \sum_{k=1}^{N}\frac{P_{k}^{2}+\omega_{k}^{2}Q_{k}^{2}}{2}+\frac{\alpha}{3}\sum_{k,j,l=1}^{N}\hspace{-5pt}A_{k,j,l}Q_{k}Q_{j}Q_{l}
\end{equation}
and
\begin{equation}\label{modalbham}
\hspace{-5pt}H_{\beta}(\boldsymbol{Q},\boldsymbol{P}) = \sum_{k=1}^{N}\frac{P_{k}^{2}+\omega_{k}^{2}Q_{k}^{2}}{2}+\frac{\beta}{4}\hspace{-8pt}\sum_{\hspace{6pt}k,j,l,m=1}^{N}\hspace{-13pt}B_{k,j,l,m}Q_{k}Q_{j}Q_{l}Q_{m}
\end{equation}
where the normal mode frequencies are
\begin{equation}
\omega_{k} = 2 \sin\left(\frac{k \pi}{2(N+1)}\right).
\end{equation}
The coupling constants $A_{k,j,l}$ and $B_{k,j,l,m}$ are given by \cite{bivins,sholl}

\begin{equation}
A_{k,j,l} = \frac{\omega_{k}\omega_{j}\omega_{l}}{\sqrt{2(N+1)}}\sum_{\pm}\left(\delta_{k,\pm j\pm l}-\delta_{k\pm j\pm l,2(N+1)}\right)
\end{equation}

\begin{equation}
B_{k,j,l,m} \hspace{-2pt}=\hspace{-2pt} \frac{\omega_{k}\omega_{j}\omega_{l}\omega_{m}}{2(N+1)}\hspace{-2pt}\sum_{\pm}\hspace{-2pt}\left(\delta_{k,\pm j\pm l\pm m}\hspace{-2pt}-\hspace{-2pt}\delta_{k\pm j\pm l\pm m,\pm 2(N+1)}\right)
\end{equation}
where the sums $\sum_{\pm}$ are over all combination of plus and minus signs among the $\pm$ symbols and $\delta_{j,l}$ is the Kronecker delta function.

We follow Ref. \cite{De Luca} and rescale the normal mode coordinate and momentum pairs in equation \ref{modalaham} by $(\bm{Q},\bm{P})\rightarrow(\bm{Q}/\alpha,\bm{P}/\alpha)$ and in equation \ref{modalbham} by $(\bm{Q},\bm{P})\rightarrow(\bm{Q}/\sqrt{\beta},\bm{P}/\sqrt{\beta})$. Letting $E$ represent energy, this leads to
\begin{equation}
H_{\alpha=1}(\bm{Q},\bm{P}) = \alpha^{2}E
\end{equation}
\begin{equation}
H_{\beta=1}(\bm{Q},\bm{P}) = \beta E
\end{equation}
which allows us to investigate results as functions of the parameters $E\alpha^{2}$ and $E\beta$, respectively.

We use the symplectic $SABA_{2}C$ integrator introduced in Ref.  \cite{Laskar and Robutel}. This integrator is reviewed and applied to FPUT lattices in the appendix of our paper. Also in the appendix, we describe how accurately the $SABA_{2}C$ integrator conserves energy at different time step sizes and show the relative energy error in the return to the initial state when our calculations are time-reversed.  

\section{Revisiting previous results}\label{revist}

To the best of our knowledge, there have yet to be studies on HoRs greater than 2nd order, and other than previously mentioned references \cite{TandM,drago and ridella,henry and sholl}, there appears to be nothing else in the literature focusing on 2oRs in FPUT lattices. Therefore, we will briefly revisit these limited previous results to set the background and context for our study.

The past studies on 2oRs all considered fixed amplitude initial conditions $q_{n}(0) = \mathcal{A}\sin\left(n\pi/(N+1)\right)$ and $p_{n}(0)=0$, which correspond to initial energies
\begin{equation}
E_{\alpha} = \mathcal{A}^{2}(N+1)\sin^{2}\left(\frac{\pi}{2(N+1)}\right)
\end{equation}
\begin{equation}
E_{\beta} = E_{\alpha}+\frac{3\mathcal{A}^{4}\beta(N+1)}{2}\sin^{4}\left(\frac{\pi}{2(N+1)}\right)
\end{equation}

Tuck and Menzel \cite{TandM} and Drago and Ridella \cite{drago and ridella} considered initial conditions with fixed amplitude $\mathcal{A}=1$. Our Fig. \ref{arecurrences}b is a replication of Tuck and Menzel's results shown in Fig. 2 of their paper. As Drago and Ridella also found, Tuck and Menzel's results for the $\alpha$-FPUT lattice are quantitatively reproducible. The time for the first 2oR in Fig. \ref{arecurrences}b, $T=24875$, is only 0.821\% shorter than the time found by Tuck and Menzel, $T=25081$, which is well within acceptable numerical error. Regarding Tuck and Menzel's results for the $\beta$-FPUT lattice, 
\begin{figure}[t]
	\includegraphics[width=.48\textwidth]{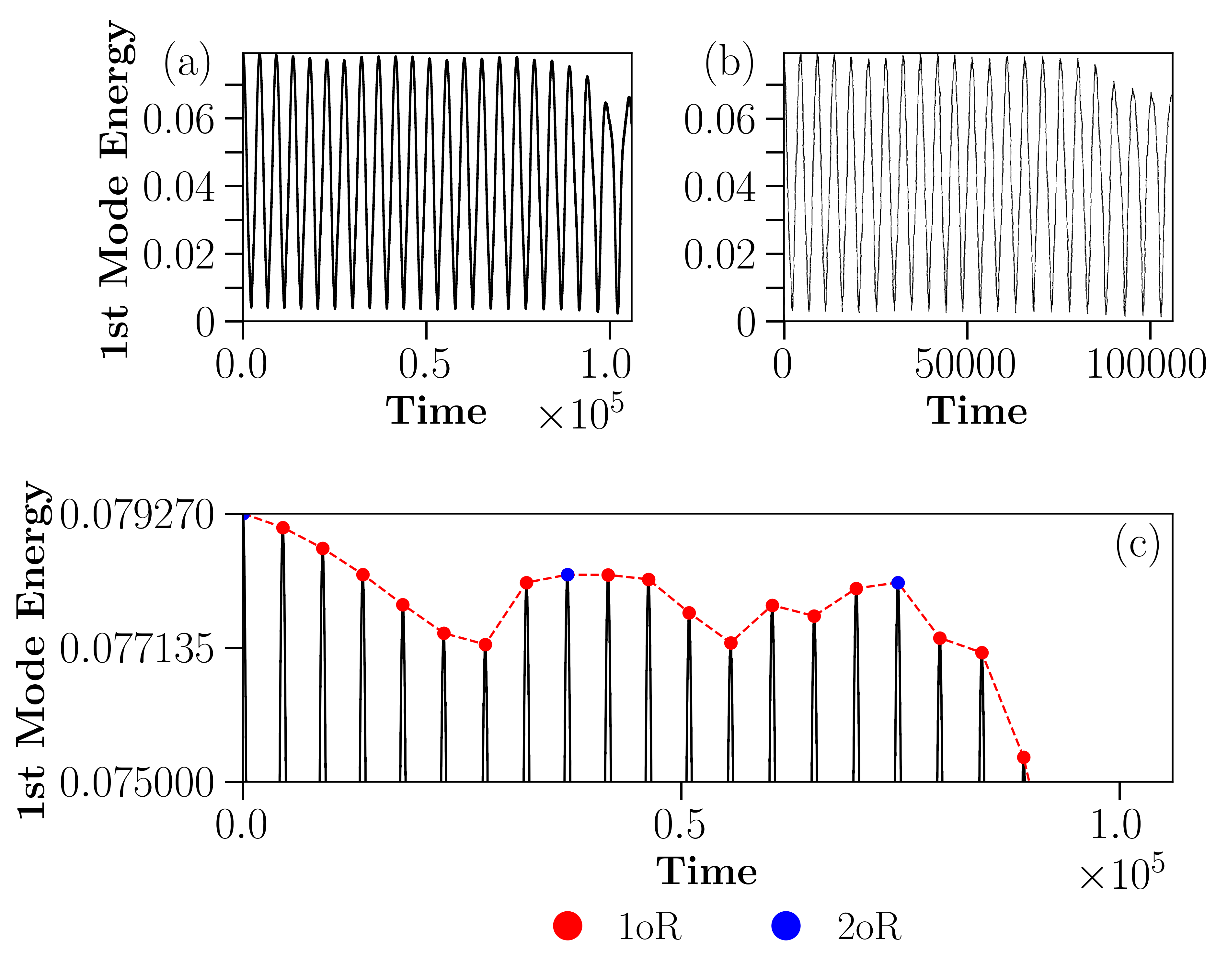}
	\caption{Using the same parameters as Tuck and Menzel \cite{TandM} ($N=31$, $\beta=8$, $E=0.07927$), (a) shows our results (with a time step size of $\tau=10^{-3}$) and (b) shows Drago and Ridella's results \cite{drago and ridella} at the lowest time step size they considered ($\tau=\sqrt{2}/400$). (c) Zooms into our results to show subtle 2oRs.}  
	\label{tmbrecurrences}
\end{figure}
Fig. \ref{tmbrecurrences} shows (a) our and (b) Drago and Ridella's results proceed using our graphics. Like Drago and Ridella, we found the time step size to influence, more than normal, the SRs thus making Tuck and Menzel's published results a numerical artifact. There remains the question of whether 2oRs exist in the $\beta$ lattice. To show that they do indeed exist, we had to use a time step size of $\tau=10^{-3}$ in order to see the dynamics accurately time-reversible. Comparing our replication (Fig. \ref{tmbrecurrences}a) to theirs (Fig. \ref{tmbrecurrences}b), it seems that our results agree. However, Fig. \ref{tmbrecurrences}c zooms into a much finer energy regime and reveals subtle 2oRs in the time scale considered. Careful inspection of Drago and Ridella's results also shows these extremely subtle 2oRs. The  requirement of having a very small time step size to correctly observe 2oRs is not always present in the $\beta$-FPUT model. We find that for both FPUT-lattices, large values of $E\beta$ and $E\alpha^{2}$ require a small time step size in order to observe SRs. This suggests that the system is strongly chaotic in these regions and the time step size is a result of an imperfect computer and integrator.

Sholl and Henry \cite{henry and sholl} used initial conditions in normal mode space $Q_{k}(0)=\delta_{k,1}$ and $P_{k}(0)=0$, which translates to a fixed amplitude $\mathcal{A}=\sqrt{2/(N+1)}$ in coordinate space. They were able to clearly see 2oRs for the $\alpha$-FPUT lattice and derived an expression for the period for general $N$. While this expression does not agree well with their (or our) numerical results, going to higher order in their perturbation scheme would presumably improve its accuracy. However, for the $\beta$-FPUT lattice, while they searched all $N<20$, they observed 2oRs only for $N=7$, due to a resonance unique to this value of $N$. Using their initial conditions we too only see a 2oRs for $N=7$. However, changing the energy, and thus the amplitude of the initial conditions, we are able to detect 2oRs for $N<20$. The energy values at which we were able to observe 2oRs were often ten times greater than the energy ranges considered by Sholl and Henry due to their fixed amplitude initial conditions. This was not surprising as at the energy values considered by Sholl and Henry the recurrences themselves were very subtle, suggesting near-integrable behavior where one would not expect to observe HoRs..

\section{Existence of Higher-Order Recurrences}\label{HoR exists}

\begin{figure}[t]
	\includegraphics[width=.48\textwidth]{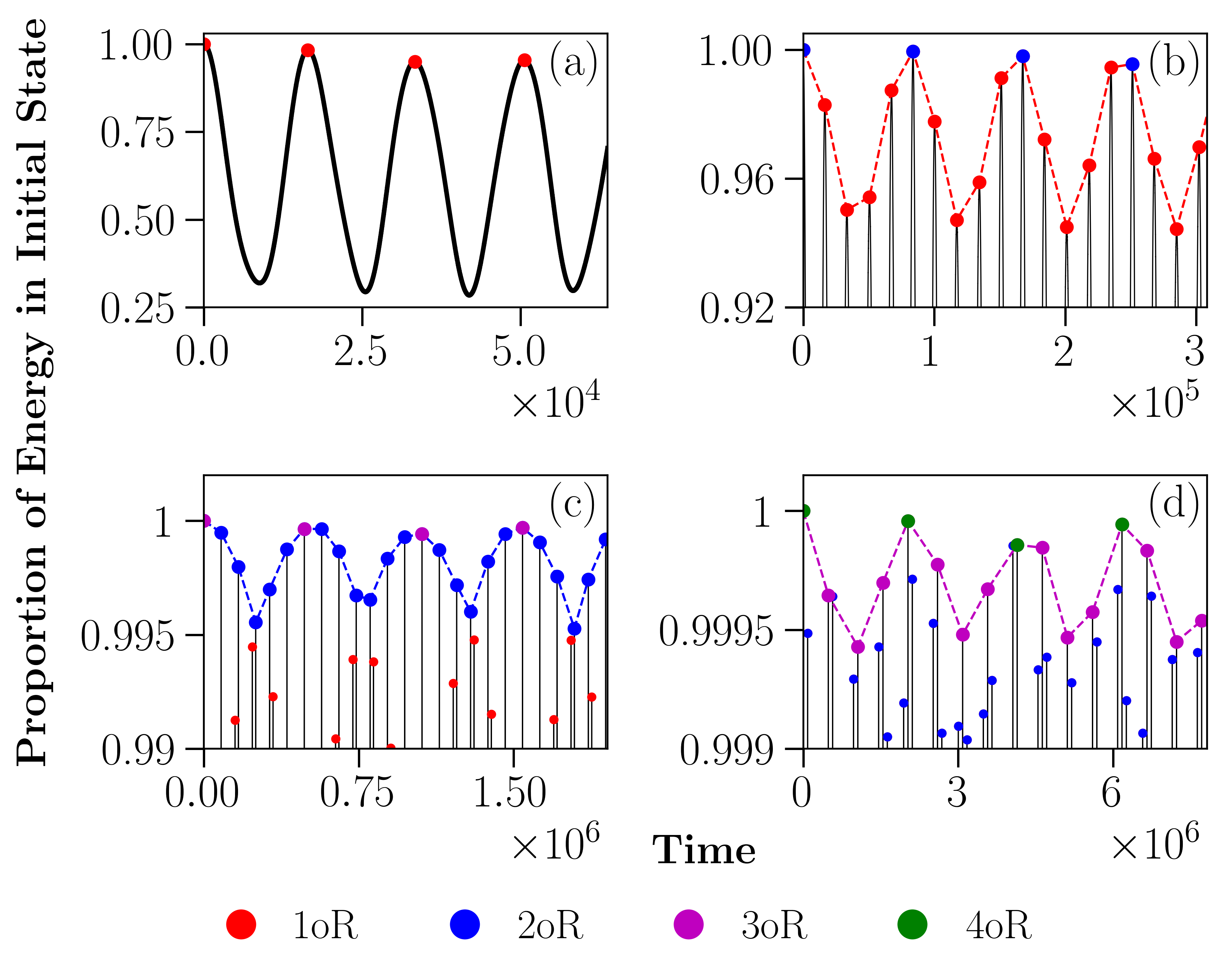}
	
	\caption{The proportion of energy in the initial state is shown to examine HoRs in the $\alpha$-FPUT lattice ($E\alpha^{2}=4.35\times 10^{-4}$ and $N=31$). (a) 1oRs ($\langle T_{R} \rangle\sim 1.68\times10^{4}$) in red, (b) 2oRs ($\langle T_{2\text{oR}} \rangle\sim 8.12\times10^{4}$) in blue, (c) 3oRs ($\langle T_{3\text{oR}} \rangle\sim 5.13\times10^{5}$) in purple, and (d) 4oRs ($\langle T_{4\text{oR}} \rangle\sim 2.05\times10^{6}$) in green.} 
	
	\label{aHoR_existence}
\end{figure}

\begin{figure}[t]
	\includegraphics[width=.48\textwidth]{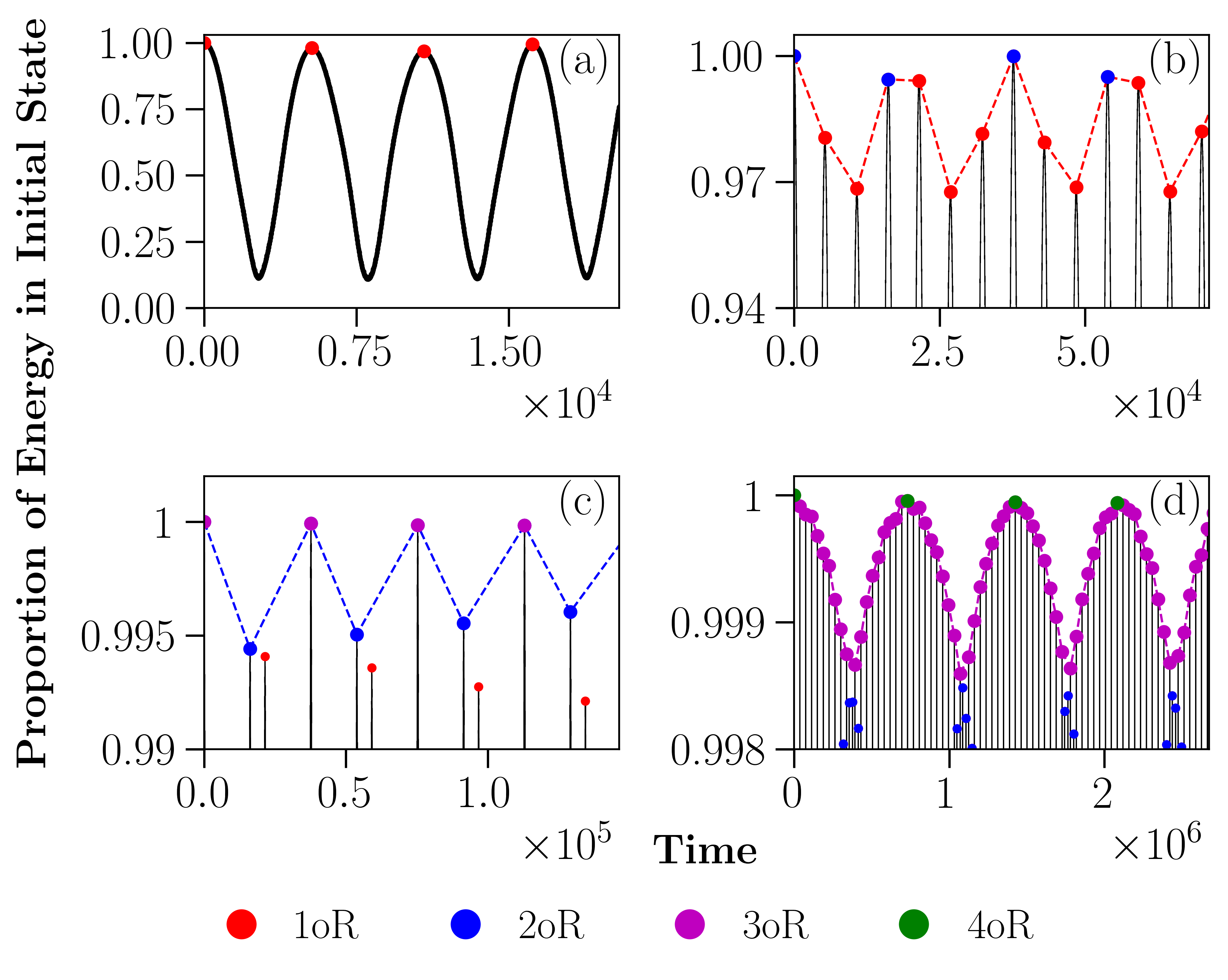}
	
	\caption{The proportion of energy in the initial state is shown to examine HoR in the $\beta$-FPUT lattice ($E\beta=0.4589$ and $N=31$). (a) 1oR ($\langle T_{R} \rangle\sim 5.38\times10^{3}$) in red, (b) 2oR ($\langle T_{2oR} \rangle\sim 1.88\times10^{4}$) in blue, (c) 3oR ($\langle T_{3oR} \rangle\sim 3.85\times10^{4}$) in purple, and (d) 4oR ($\langle T_{4oR} \rangle\sim 6.94\times10^{5}$) in green.} 
	
	\label{bHoR_existence}
\end{figure}

We demonstrate the existence of HoRs for the $\alpha$-FPUT lattice in Fig. \ref{aHoR_existence} and for the $\beta$-FPUT lattice in Fig. \ref{bHoR_existence}, which both show the proportion of energy in the initial state (fundamental mode) in black. Note the different energy and time scales considered in the different parts of Figs. \ref{aHoR_existence} and \ref{bHoR_existence}. The points represent recurrences and the dashed lines connecting the points are meant to emphasize the structured modulation among the recurrences. Figs.  \ref{aHoR_existence}a and \ref{bHoR_existence}a show the 1oRs in red. Figs. \ref{aHoR_existence}b and \ref{bHoR_existence}b show the modulation of the 1oRs and reveal 2oRs in blue. Figs. \ref{aHoR_existence}c, \ref{bHoR_existence}c, \ref{aHoR_existence}d, and \ref{bHoR_existence}d continue this and show 3oRs in purple and 4oRs in green.


It is important to note the subtly of most HoRs. Sholl and Henry \cite{henry and sholl} mentioned that while the shifted-frequency perturbation method provides the necessary tools to investigate any other HoRs, their modulations would be extremely small. This is in agreement with our numerical results. Furthermore, for smaller values of $E\alpha^{2}$ and $E\beta$ these HoRs became increasingly subtle, as one would expect, since the lattices are nearing their integrable limits. Because these modulations can be small, we have confirmed all of our results by verifying the time-reversal symmetry of microscopic dynamical equations. This was done by running the simulation forward in time from $t=0$ to $t=\mathcal{T}$ and then reversing time and running the simulation backward in time from $t=\mathcal{T}$ back to $t=0$. The behavior forward in time and backward in time matched very closely qualitatively; for a quantitative comparison,  the relative error of the energy returned to the initial state is shown in detail in the appendix.

\begin{figure}[t]
	\includegraphics[width=.48\textwidth]{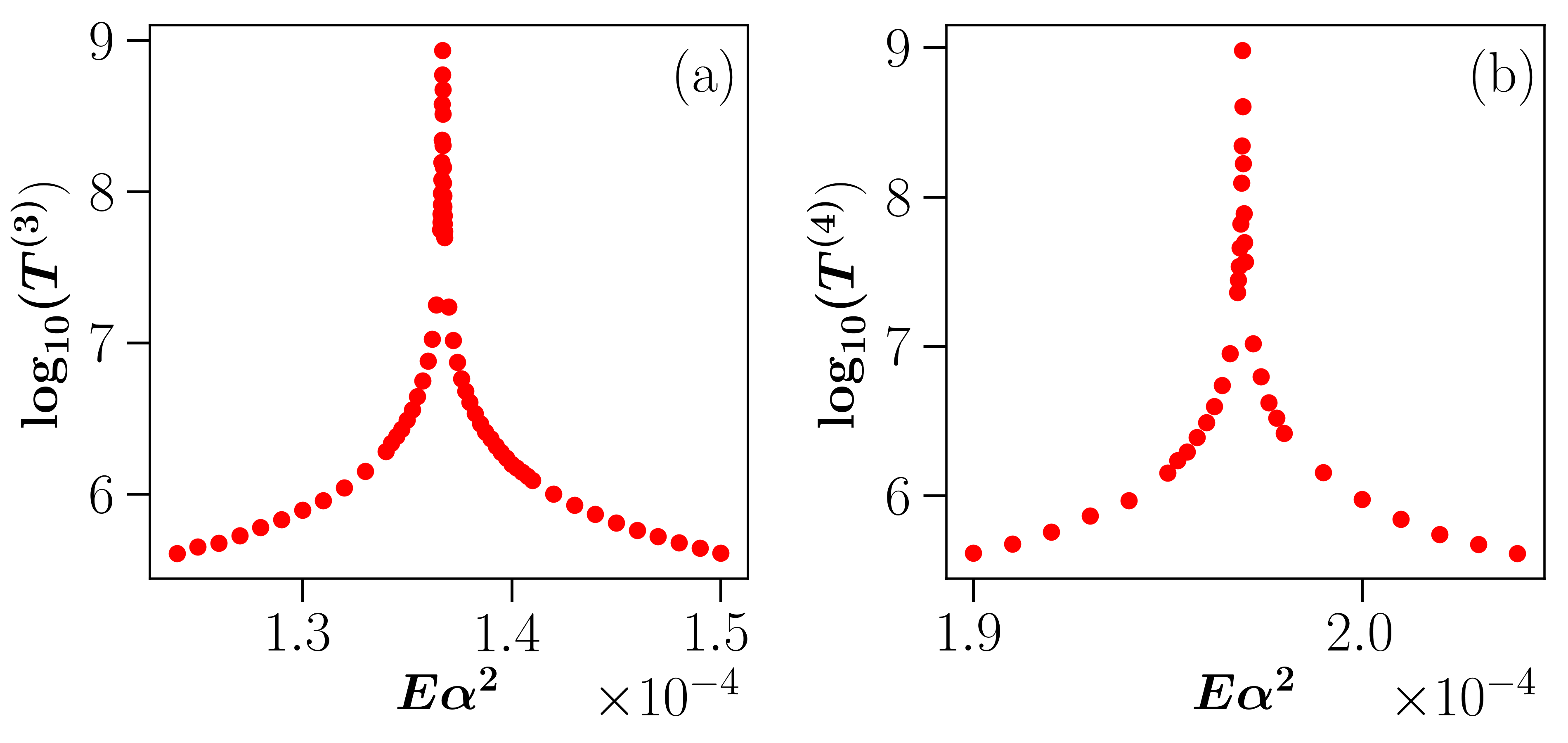}
	
	\caption{Shows apparent singularities in the (a) 3-period and (b) 4-period for the $\alpha$-FPUT lattice with N=31.}
	\label{a_HoR_singularities}
	\bigbreak
	\includegraphics[width=.48\textwidth]{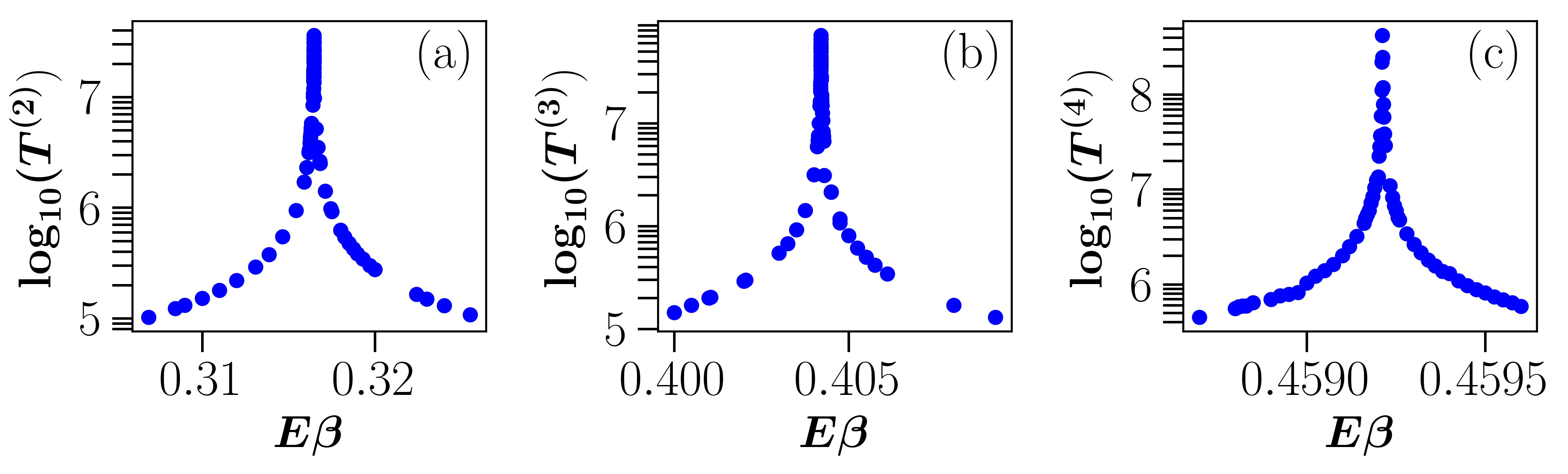}
	\caption{Shows apparent singularities in the (a) 2-period and (b) 3-period, and (c) 4-period for the $\beta$-FPUT lattice with N=31.}
	\label{b_HoR_singularities}
\end{figure}

\section{Scaling of Higher-Order Recurrence Times}\label{HoR}
We preformed numerical experiments with $N=31$ to see how the periods of HoRs depend on $E\alpha^{2}$ and $E\beta$. The $n$oR period, $\langle T_{n\text{oR}} \rangle$, was obtained by averaging over the time differences between many neighboring HoRs
\begin{equation}
\langle T_{n\text{oR}} \rangle = \frac{T_{n\text{oR}}^{\text{1st}}}{w}+\frac{1}{w}\sum_{j=2}^{w}(T_{n\text{oR}}^{j\text{th}}-T_{n\text{oR}}^{(j-1)\text{th}})=\frac{T_{n\text{oR}}^{w\text{th}}}{w},
\end{equation}
where $w$ is an integer chosen such that increasing $w$ further would not change $\langle T_{n\text{oR}} \rangle$. This was important because the $n$oRs can only occur in integer numbers of the (n-1)oRs; this causes fluctuation in the $n$oR times based on which $n$oR one chooses. For example, in Fig. \ref{aHoR_existence}c, the 3oR occurs on the 6th, then 13th, then 19th 2oR, instead of what one might naively expect, the 6th, 12th, then 18th. Therefore, the $T_{3\text{oR}}^{\text{1st}}$ will be about $\langle T_{2\text{oR}} \rangle$ less than the $T_{3\text{oR}}^{\text{2nd}}$ and $T_{3\text{oR}}^{\text{3rd}}$ and hence taking the average makes more sense. 

\begin{figure}[t]
	\includegraphics[width=.48\textwidth]{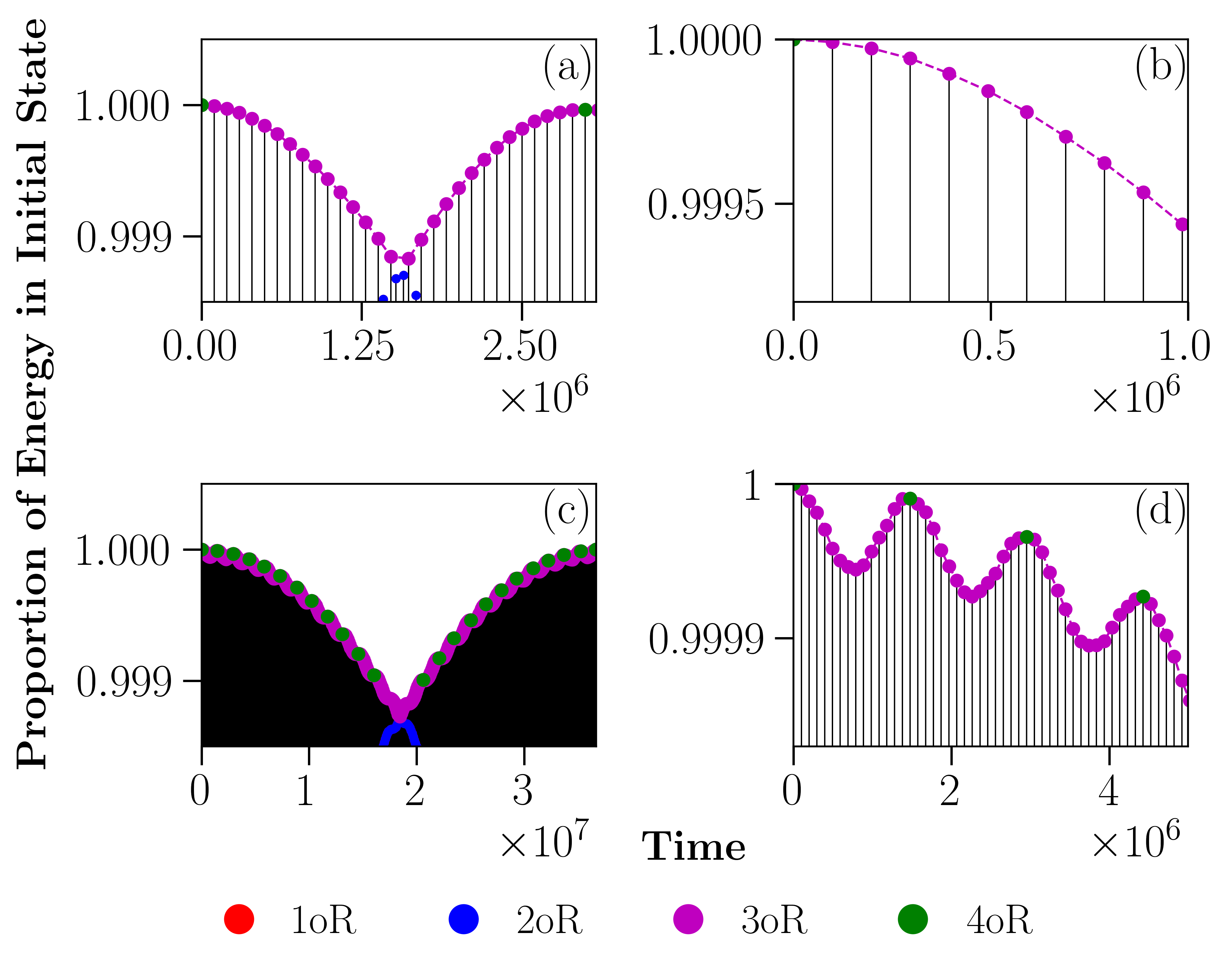}
	\caption{The formation of nested HoRs are shown in the $\alpha$-FPUT lattice approaching an apparent singularity, going from (a-b)  $E\alpha^{2}=0.000196$ and $\langle T^{(4)} \rangle=3.08\times 10^{6}$ to (c-d) $E\alpha^{2}=0.000197$ and $\langle T^{(4)}\rangle=3.67\times 10^{7}$. (b) and (d) are zoomed in figures of (a) and (d), respectively, to highlight this phenomena.}
	\label{aHoR_nested_HoR}
	\bigbreak
	\includegraphics[width=.48\textwidth]{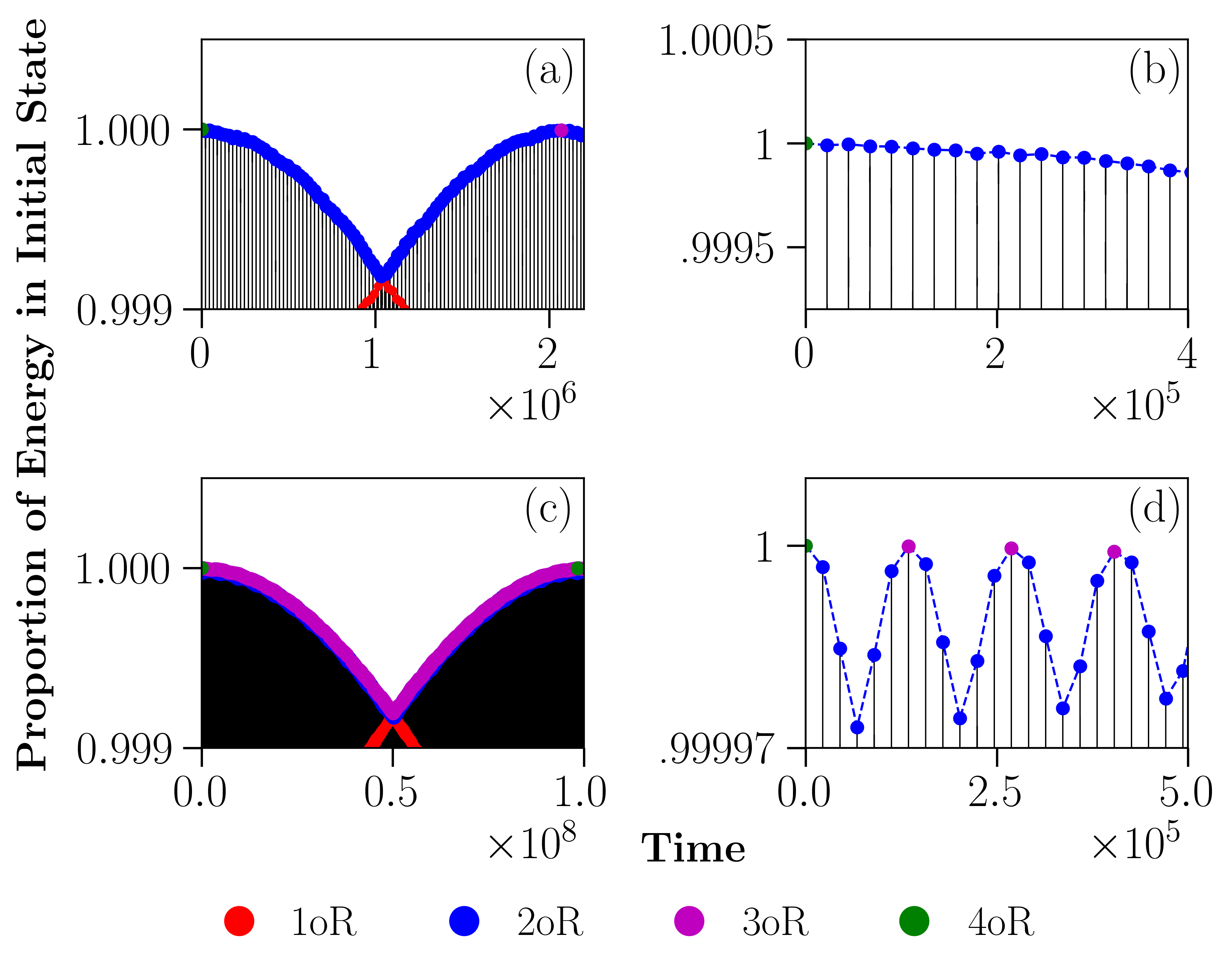}
	
	\caption{The formation of nested HoRs is shown in the $\beta$-FPUT lattice approaching an apparent singularity, going from (a-b) $E\beta=0.27315$ and $\langle T^{(3)} \rangle=2.07\times 10^{6}$ to (c-d) $E\beta=0.27301$ and $\langle T^{(3)}\rangle=9.84\times 10^{7}$. (b) and (d) are zoomed in figures of (a) and (d), respectively, to highlight this phenomena.}
	\label{bHoR_nested_HoR}
\end{figure}

We have found that, contrary to the original FPUT recurrences, the scaling of HoRs times is nontrivial due to the existence of apparent singularities in the HoR periods as functions of the parameters. Examples of these singularities are shown in Figs. \ref{a_HoR_singularities} and \ref{b_HoR_singularities}. There is another layer of complexity in these divergences due to the formation of ``new'' HoRs: that is, at long periods, the HoRs change their order due to the appearance of new HoRs with shorter periods and smaller fluctuations than the HoRs first considered. We call these "nested" HoRs and give examples Figs.  \ref{aHoR_nested_HoR} and \ref{bHoR_nested_HoR}. Therefore, the original HoRs will "increase"  their order as these new HoRs appear.

To categorize these HoRs periods, we will denote by $T^{(n)}$ the period an $n$-period HoR  if it initially, before any new HoRs form through the scenario mentioned above, is the period of $n$oRs. For example, if $E\beta$ approaches a 2-period singularity from the left, the period will initially describe 2oRs. However, as $E\beta$ gets closer to the singularities center, eventually new 2oRs will form with periods much shorter than the period of the original 2oRs, which are now 3oRs. If $\gamma$ is the critical value of $E\beta$ at which the new HoRs form, then
\begin{equation}
\lim_{E\beta \to \gamma^{-}}T_{2oR}(E\beta)=\lim_{E\beta \to \gamma^{+}}T_{3oR}(E\beta) = T^{(2)}(E\beta=\gamma).
\end{equation}

Looking first at the $\alpha$-FPUT lattice, we see in Fig. \ref{a_HoR_singularities}a an apparent singularity in the 3-period centered about $E\alpha^{2}\sim 1.366\times 10^{-4}$ and Fig. \ref{a_HoR_singularities}b shows an apparent singularity in the 4-period centered about $E\alpha^{2}\sim 1.97 \times 10^{-4}$. It is interesting to note that for $N=31$ and $10^{-9}<E\alpha^{2}<10^{-2}$ while we observed multiple singularities in the 3 and 4-periods, we did not observe any singularities in the 2-period. Comparing Figs \ref{a_HoR_singularities}a and \ref{a_HoR_singularities}b, it appears that for higher order recurrence periods, the singularities are narrower. Therefore, the higher-order the recurrence, the smaller the $E\alpha^{2}$ regime in which it exists. 

As noted before, when an $n$-period is large, one sees nested HoRs form, with orders less than $n$ and modulations less than the $n$oRs modulations. Fig. \ref{aHoR_nested_HoR} shows an example of these HoRs forming as a 4-period becomes large. The color coding used in Figs. \ref{aHoR_existence} and \ref{bHoR_existence} still applies. Fig. \ref{aHoR_nested_HoR}a shows the 4-period, which describes 4oRs as seen in Fig. \ref{aHoR_nested_HoR}b which zooms in on Fig. \ref{aHoR_nested_HoR}a. Fig. \ref{aHoR_nested_HoR}c shows the same 4-period but at a slightly different value of $E\alpha^{2}$, which is closer to the center of the singularity. Fig. \ref{aHoR_nested_HoR}d once again zooms in and shows that the 3oRs have picked up a new modulation which was not present at the lower 4-period value, and therefore, the 4-period now describes 5oRs. 

As for the $\beta$-FPUT lattice, Fig. \ref{b_HoR_singularities} shows apparent singularities of the 2-period, 3-period, and 4-period. In contrast to the $\alpha$-FPUT lattice, for $N=31$, the $\beta$-FPUT lattice exhibits apparent singularities in its 2-period. Other than this, the same trends for the $\alpha$-FPUT lattice $n$-periods exist in the $\beta$-FPUT lattice $n$-periods. One such trend is the formation of nested HoRs at large $n$-periods, which is shown in Fig. \ref{bHoR_nested_HoR}. The same phenomenon as described above for the $\alpha$-FPUT lattice occurs, but now it is a 3-period which has a singularity.

With respect to the height of these apparent singularities, within the limits of our computational resources, it does appear that at precisely right value of $E\alpha^{2}$ or $E\beta$, the period may indeed diverge to infinity, as we have not observed a maximum to the peaks. As we shall discuss in more detail below, this strongly suggests that at certain energies some exact nonlinear resonance occurs, with a vanishing denominator leading to an infinite period.

In our studies of these singularities, the omission of $n$-periods other than 4 in both lattices is because we have not observed these periods. However, this does not suggest that none exist, as higher ordered periods exist in ever finer scaling regimes, which makes it harder to find the precise values of energy, alpha, and beta to observe them. We have chosen not to pursue these more detailed searches at this time. 

Sholl and Henry's theoretical expression for the recurrence and super-recurrence periods involves resonances between nonlinear ("shifted") frequencies and thus take the form
\begin{equation}\label{HS_T}
T=\frac{1}{\sum_{k} c_{k}\Omega_{k}},
\end{equation}
where $\Omega_{k}$ is a nonlinear frequency and $c_{k}$ is some integer. The nonlinear frequencies are defined by the perturbation scheme as $\Omega_{k}^{2}=\omega_{k}^{2}+\sum_{j=1}\mu_{k,j}\hspace{1pt}\chi^{j}$ where $\chi$ is the nonlinear parameter (ie. $\alpha$ or $\beta$) and $\mu_{k,j}$ are what they call ``frequency corrections", defined to eliminate secular terms. An apparent singularity in the theoretically derived 2-period for $N=7$ in the $\beta$-FPUT lattice is shown in figure 7 of Sholl and Henry's paper but they did not discuss this. Their 2-period expression, which is dependent on $\beta$ only, is given by 
\begin{align}
\begin{split}
T^{(2)}&=\frac{2\pi}{5\Omega_{1}-\Omega_{7}} \\
&=\frac{2\pi}{5\sqrt{\omega_{1}^{2}+\mu_{1,1}\beta+\mu_{1,2}\beta^{2}+\mu_{1,3}\beta^{3}}-\omega_{7}}
\end{split}
\end{align}
Therefore, it is reasonable to conclude that the apparent singularities  we have observed are also caused by near resonances between nonlinear normal modes, which a blow up in the period due to a small denominator in expressions like equation \ref{HS_T}.

Before turning to the next section on the breakdown of the 2oRs and how this relates to the approach to equilibrium, we will end this section with a qualitative summary of the rather complicated behavior of the model dynamics in the different energy regimes relevant to recurrences. For very small values of $E\alpha^{2}$ and $E\beta$, when the lattices are approximately integrable, not only do the original FPUT recurrences become increasingly subtle, but the HoRs do as well. In these low energy regimes, we have not observed any apparent singularities in the HoR periods. As $E\alpha^{2}$ and $E\beta$ increase and the original FPUT recurrences become more noticeable, we observe the behavior discussed above: namely, HoRs (which amount to periodic modulations of recurrences) exhibit singularities in which their periods seem to go to infinity at very specific energies, etc. The 2oRs in the $\alpha$-FPUT lattice appear to always exist with periods decreasing with increasing energy, and they never exhibit singularities in their periods. In the $\beta$-FPUT lattice, 2oRs do not always exist. When they do exist, their behavior is non-monotonic. In particular, as shown above, for values of $E\beta$ where there are no 2oRs, slightly decreasing or increasing $E\beta$ leads to $2oR$s with periods behaving in a singular manner, first increasing (apparently) to infinity and then deceasing with changing energy. This is the same behavior seen in the third and higher order recurrences of both lattices: again they behave non-monotonically with energy and exhibit singularities in their periods. Between the energy values centering on the singularities, they do not exist. At still higher values of the energy parameters, above some critical energy, the periods of the 2oRs in both lattices start to increase with energy, at a much slower rate than at a singularity, until they begin to breakdown in their particular way depending on the lattice. In the ensuing section, we will study this breakdown in detail in the two lattices.

\section{The Breakdown of 2nd-Order Recurrences and Thermalization}\label{breakdown}

FPUT recurrences and HoRs are a defining feature that explains the lack of rapid thermalization in FPUT-lattices at low energies. Therefore, one may expect that the breakdown (or lack of formation) of HoRs will indicate that the system is approaching equilibrium. Increasing $E\alpha^{2}$ and $E\beta$  to large values enough allows one to observe the breakdown (or lack of formation) of 2oRs due to the changing of thermalization timescales as the system passes the ``strong stochastic threshold'' \cite{pettini,berch, Dan camp flach}. Surprisingly, we find that while this clear intuition--that is, of HoRs breaking down being affiliated with the lattice approaching equilibrium--is indeed the case for the $\beta$-FPUT lattice, it is not the case for the $\alpha$-FPUT lattice.

\begin{figure}[t!]
	\includegraphics[width=.47\textwidth]{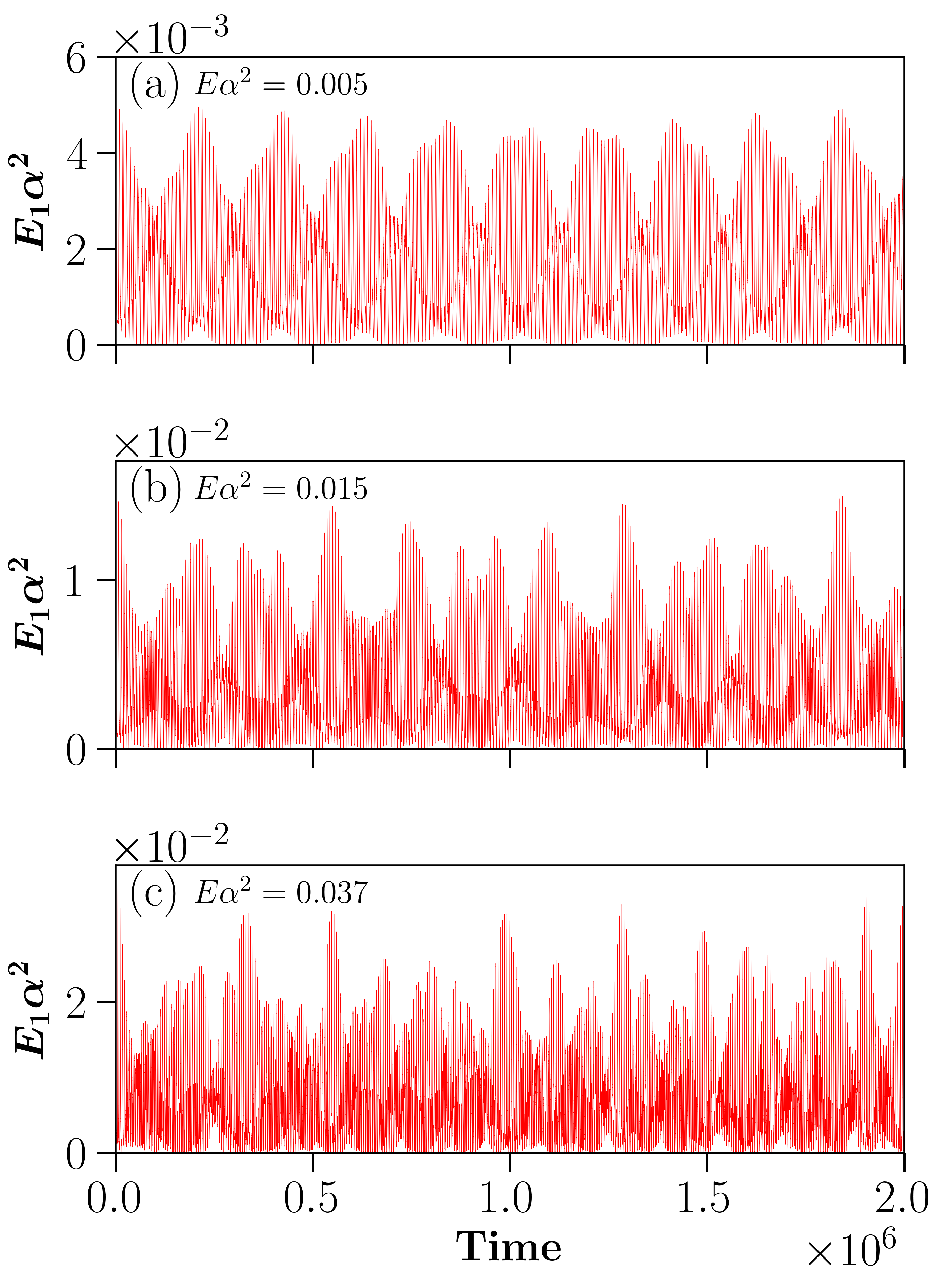}
	
	\caption{The break down of 2oRs for the $\alpha$-FPUT lattice with (a) $E\alpha^{2}=0.005$, (b) $E\alpha^{2}=0.015$, (c) $E\alpha^{2}=0.037$. Increasing $E\alpha^{2}$ cause the 2oRs to lose their shape on a very short time scale and thus never truly form.} 
	
	\label{aSRbreakdown}
	\bigbreak
	\includegraphics[width=.47\textwidth]{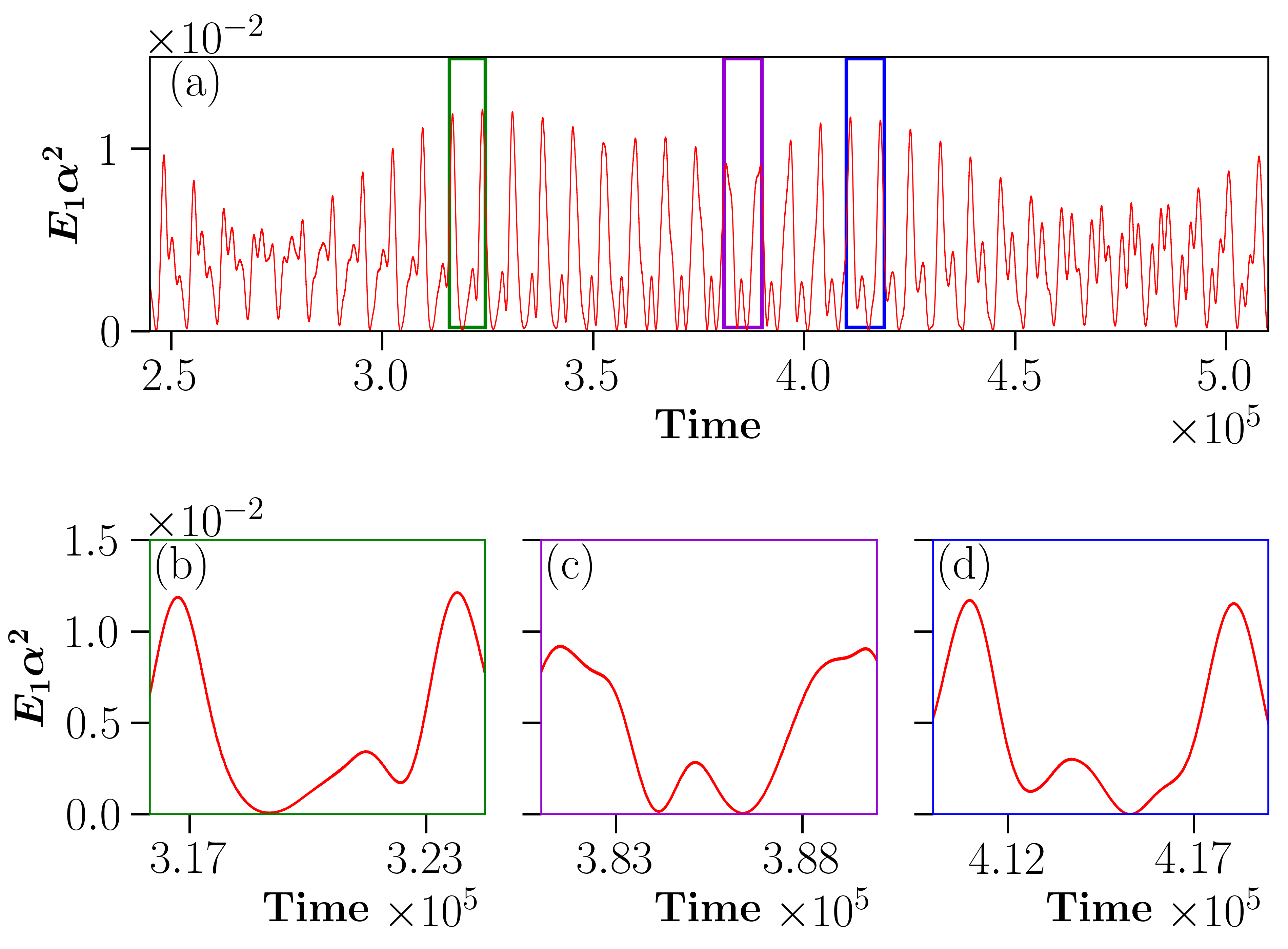}
	
	\caption{Zooms into Fig. \ref{aSRbreakdown}b, where $E\alpha^{2}=0.015$, to see what is causing the deformation of the 2oRs. (a) Shows this zoomed region, and (b)-(d) show further zoomed regions on time. } 
	
	\label{aSRzoom}
\end{figure}

In the $\alpha$-FPUT lattice, the 2oRs break down due to a degradation of their structure which becomes more apparent as $E\alpha^{2}$ increases and occurs at very short time scales, as shown in Fig.  \ref{aSRbreakdown}. Comparing the Figs. \ref{aSRbreakdown}a-c, one observes that this deformation happens slowly as one increases $E\alpha^{2}$. The cause of this deformation can be understood through Fig. \ref{aSRzoom}, which shows Fig.  \ref{aSRbreakdown}b at a zoomed in timescale. As shown in \ref{aSRzoom}a, when 2oRs in the $\alpha$ lattice breakdown, the periodic envelope dictating the modulation of FPUT recurrences loses its shape. Figs. \ref{aSRzoom}b-d show that this occurs due to a secondary FPUT recurrence, a "mini-recurrence (mR)", that has a different period than the FPUT recurrences. These mRs thus shift in time and cause some FPUT recurrences to be normal and others to fail to restore the initial energy. For $N=31$, we find that $E\alpha^{2}=0.0014$ is the threshold for the formation of mRs. For $0.0014 < E\alpha^{2}< 0.0078$, mRs arise in such a way to cause 3oRs, shown in Fig. \ref{aSRbreakdown}a, which scales like the singularities we have already discussed. But after $E\alpha^{2}=0.0078$, no other HoRs form and the breakdown of 2oRs with increasing energy truly commences. This shifting of the mRs can also be seen in Fig. \ref{aSRbreakdown}. The darkening of the graphs is due to an increase in the lines' density which is ultimately due to mRs. As $E\alpha^{2}$ continues to increase, more mRs existence and form with greater maxima as well. This process ultimately causes the 2oRs in the $\alpha$-FPUT lattice to cease forming, as the mRs ``steal'' more and more energy from the FPUT recurrences.

When the 2oRs are losing their shape with increasing $E\alpha^{2}$, it is as if their periods freeze in place and the mR grows and takes over the modal energy dynamics. This freezing in place of the 2oRs period can be seen by considering the 2oRs periods as $E\alpha^{2}$ approaches the breakdown value, which is shown in Fig. \ref{2p-brk}a. The 2oR period increases linearly until $E\alpha^{2} = 0.0014$, which happens to be the same value at which the mRs started forming. After this, the period plateaus. Interestingly, there is a dip in this plateau at $E\alpha^{2} = 0.004$. At this dip, we find a singularity in the 3-period, which is not shown but takes the same scaling form as the previous singularities shown. Therefore, we believe that this dip is due to the existence of nonlinear resonances unique to the value of $E\alpha^{2} = 0.004$. 

\begin{figure}[t]
	\includegraphics[width=.47\textwidth]{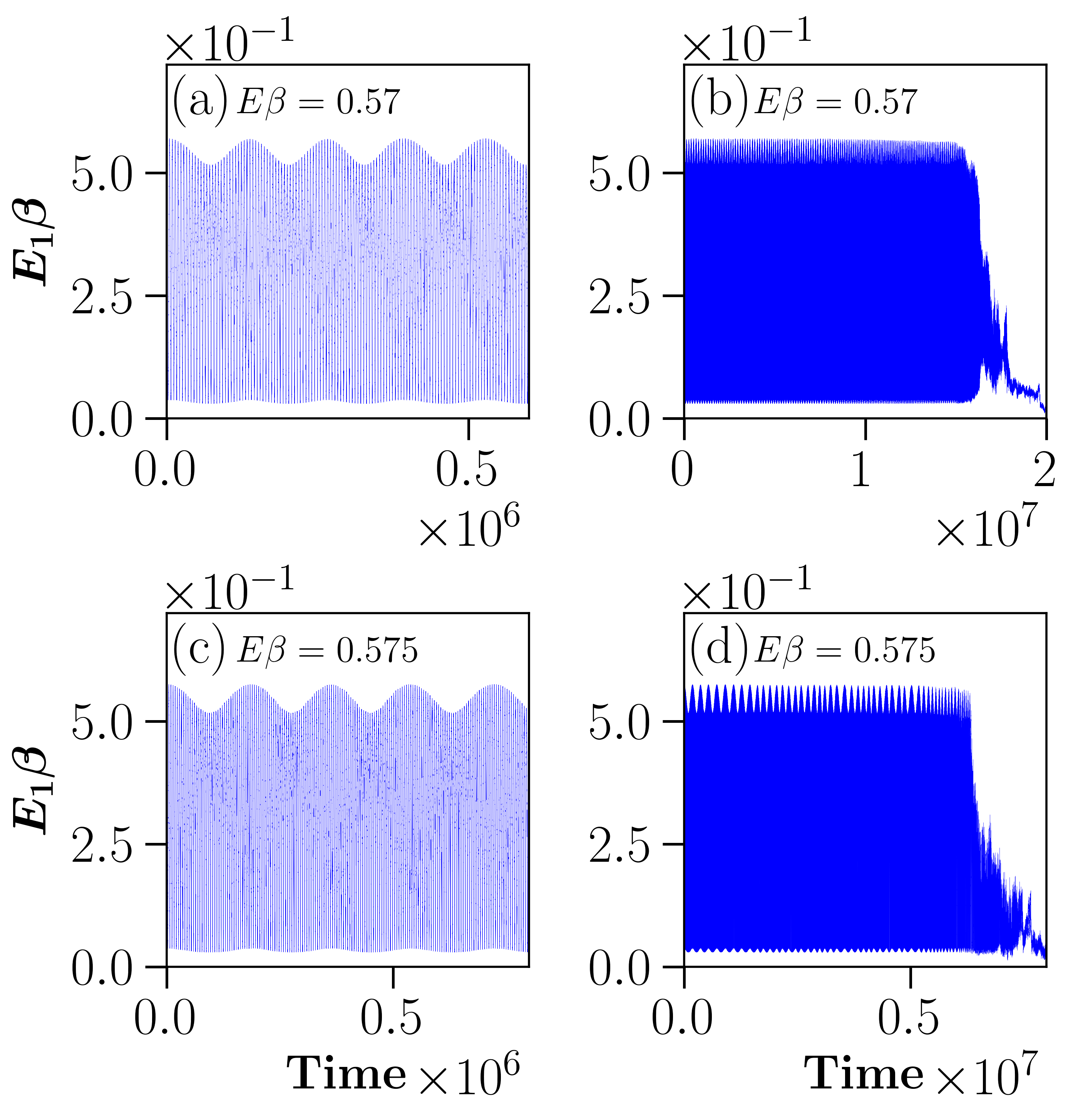}
	
	\caption{Shows breakdown of 2oRs in the $\beta$-FPUT lattice for (a-b) $E\beta=0.57$, (c-d) $E\beta=0.575$. Please note the change in time scale between parts a-b and parts c-d.} 
	
	\label{bSRbreakdown}
\end{figure}

For the $\beta$-FPUT lattice, the 2oRs do not exhibit the deformation discussed above, and no mRs are formed. Instead, the $\beta$-FPUT lattice's 2oRs breakdown abruptly, as shown in Fig.  \ref{bSRbreakdown}. Because of the time scale considered, the 2oRs appear as fast oscillations in Figs. \ref{bSRbreakdown}b and d, and thus we include parts a and c to show that these are indeed 2oRs. For $E\beta < 0.59$, the time to breakdown decreases monotonically with increasing $E\beta$, although after $E\beta = 0.59$  the monotonic behavior ceases. Interestingly, this breakdown time is quite sensitive to $E\beta$, contrary to the $\alpha$-FPUT case for which the deformation of the 2oRs deformation did not depend sensitively on $E\alpha^{2}$. 

Importantly, for Fig. \ref{bSRbreakdown} and \ref{aSRbreakdown}c, we were \textit{not} able to time-reverse the dynamics to return to the initial state;  this task lies is beyond the scope of our computation resources. The challenge in time-reversing the dynamics in these regions of $E\alpha^{2}$ and $E\beta$ results from the strongly chaotic nature of the phase space. In particular, the Lyapunov exponent in both FPUT lattices has been shown to large in both of these energy regimes \cite{pettini}. Thus, because of the large Lyapunov exponent, a very small time step size is required to time-reverse the dynamics \cite{Lozi}. Despite the lack of reversibility, we have confidence in our results because reducing the time step size further does not change the behavior of the forward time dynamics or breakdown times of  Fig. \ref{bSRbreakdown} and \ref{aSRbreakdown}c, suggesting the nature of the time evolution to be correct. 

The 2oRs that exist before they breakdown do not always scale straightforwardly with $E\beta$. Fig. \ref{2p-brk} shows the scaling of their periods as $E\beta$ and $E\alpha^{2}$ approach a regime where 2oRs breakdown on small enough time scales to be observed. There is a monotonic increase until $E\beta = 0.581$ where the period seems to vary strangely with $E\beta$ but is still similar to the nearby values. This behavior continues until $E\beta < 0.59$. We find that for $E\beta>0.59$, while the breakdown mechanism does not change, the 2oRs modulations, existence, and periods abruptly change their dependence on energy. The breakdown time fluctuates between quickly breaking down and breaking down at a time range similar to $E\beta < 0.59$ . This is the region of $E\beta$ considered by Tuck and Menzel and then Drago and Ridella.

An interesting note is that despite the differences, one similarity is that in both the $\alpha$ and $\beta$-FPUT lattices, the 2-period monotonically increases until $T^{(2)} \approx 2.1 \times 10^{5}$, as shown in Fig. \ref{2p-brk}, after which the 2oR period starts to act differently, until the 2oRs breakdown. 

\begin{figure}[t]
	\includegraphics[width=.47\textwidth]{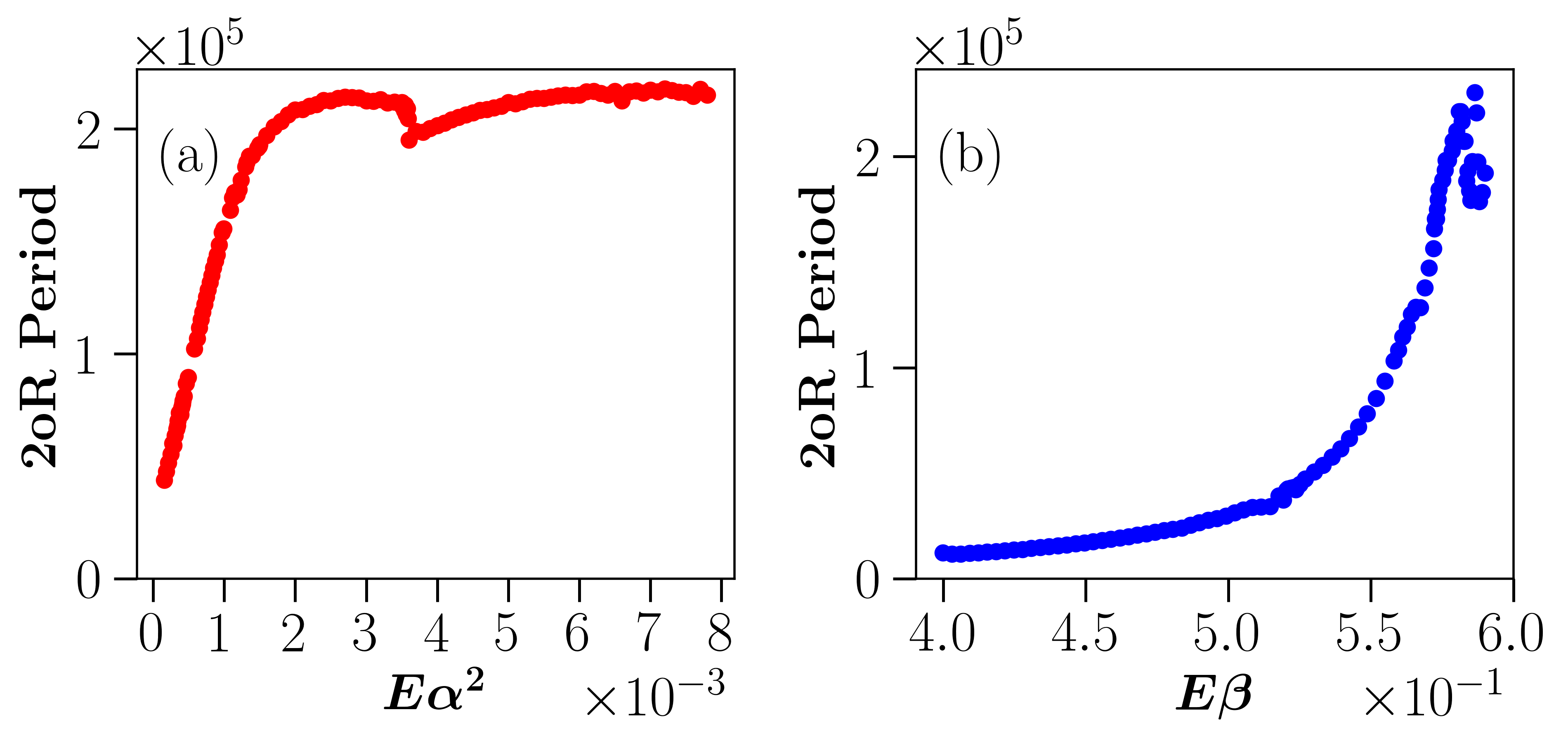}
	
	\caption{Shows the scaling of the 2-period in the (a) $\alpha$-FPUT-lattice and (b) $\beta$-FPUT lattice before 2oRs begin to breakdown. Both graphs show a region of $E\alpha^{2}$ and $E\beta$ where the 2-period is monotonically increases until changes occur when the periods are about $2.1 \times 10^{5}$.} 
	
	\label{2p-brk}
\end{figure}

A natural question to ask is how do the breakdown mechanisms of 2oRs relate to the breakdown of the apparent stationary state "metastable" state that has been widely observed \cite{Benettin}. This idea is that below a specific energy ($E/N$) threshold, there exists an apparent stationary state that causes the FPUT lattice to thermalize on a much slower timescale than above this energy threshold. This was first proposed by Fucito et al in 1982 \cite{Fucito} and was later stressed by Berchialla et al \cite{berch}. More recently, it has been interpreted in terms of \textit{q-breathers} \cite{flach et al} (which are essentially \textit{breathers} \cite{campbell et al} in mode space) and more generally in terms of \textit{q-tori} \cite{christodoulidi}. To investigate this relationship we will use spectral entropy as an indicator of equipartition and the apparent stationary state \cite{Livi,alpha yuri,lapo, Dan camp flach}. Essentially a variant of Shannon entropy, the spectral entropy is defined as
\begin{equation}
S(t) = -\sum_{k = 1}^{N}e_{k}\ln(e_{k}),
\end{equation}
where $e_{k}(t)= E_{k}(t)/\sum_{k}E_{k}(0)$ is the proportion of energy in the normal mode $k$. Spectral entropy ranges from 0 when all the system's energy is in one normal mode, to a maximum value when energy is shared equally among the normal modes. For the initial states considered, the $\alpha$-FPUT lattice this range is $0\leq S_{\alpha}(t)\leq \ln(N)$, while for the $\beta$-FPUT lattice it is $0\leq S_{\beta}(t)\leq \ln(\lceil N/2 \rceil)$. Here $\lceil$ $\rceil$ denotes the ceiling function which maps a real number to its largest, nearest integer. The maximum spectral entropy for the $\beta$-FPUT lattice is $\ln(\lceil N/2 \rceil)$ because energy can be shared only among $\lceil N/2 \rceil$ degrees of freedom, since the even modes never having energy in this lattice when the initial state is in the first mode. If the initial state were a linear combination of both even and odd normal modes, the maximum spectral entropy for the $\beta$-FPUT lattice would be the same as for the $\alpha$-FPUT lattice, $\ln({N})$

\begin{figure}[t!]
	\includegraphics[width=.47\textwidth]{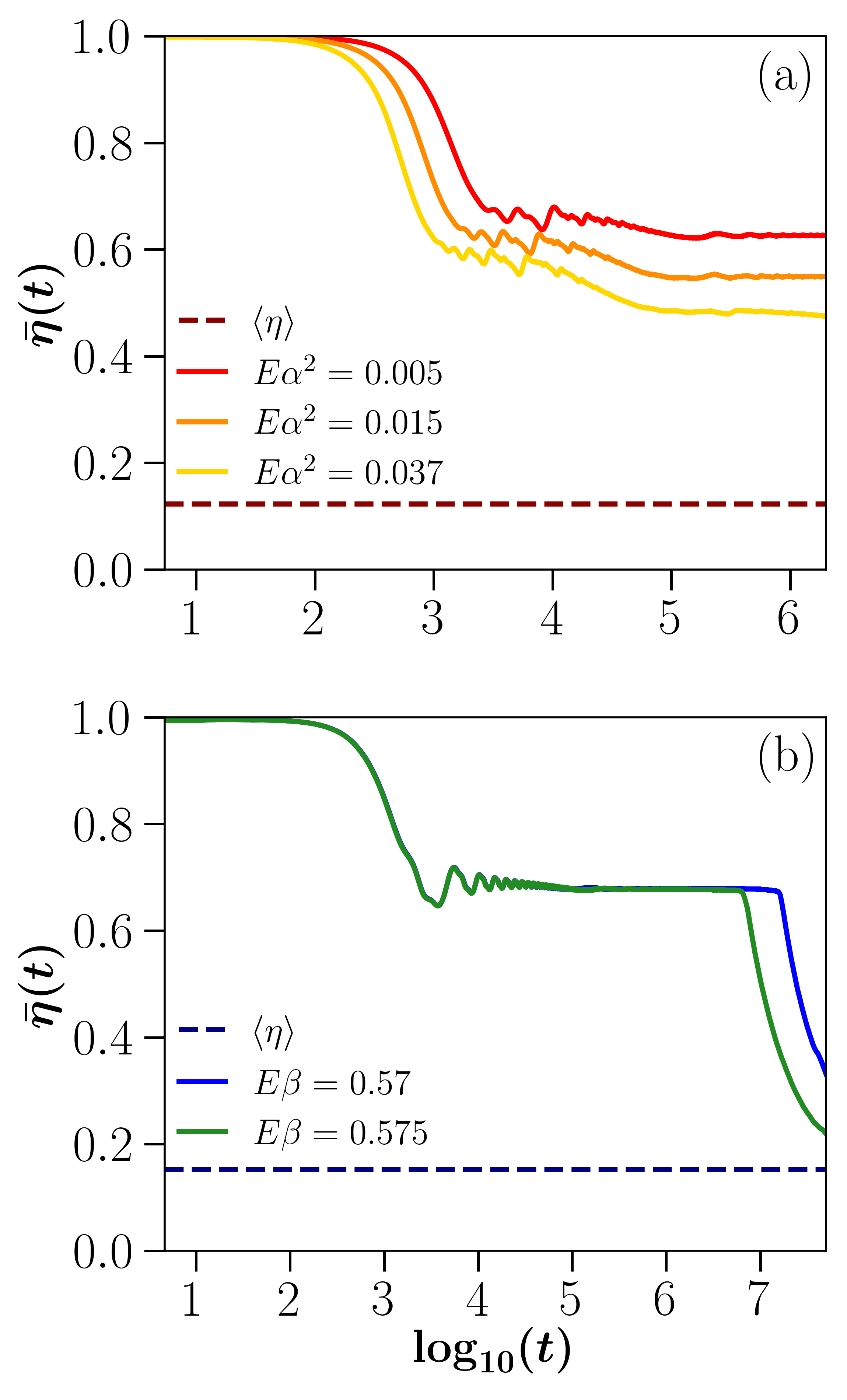}
	
	\caption{The time averaged, rescaled spectral entropy is shown for the parameters considered in Figs (a) \ref{aSRbreakdown} and (b) \ref{bSRbreakdown}. The breakdown of 2oRs in the (a) $\alpha$-FPUT lattice is not associated with the breakdown of the metastable state, while it is for the (b) $\beta$-FPUT lattice. } 
	
	\label{SR_meta}
\end{figure}

For convenience we rescale the spectral entropy so that it ranges from zero to unity and is thus independent of $N$. This is done by introducing
\begin{equation}
\eta (t) = \frac{S(t)-S_{max}}{S(0)-S_{max}}.
\end{equation}

To analyze whether the FPUT lattices has reached equipartition, we compare the time average of the normalized spectral entropy ($\bar{\eta}$) to its ensemble average ($\langle \eta \rangle$), such that
\begin{equation}
\bar{\eta}(t)=\frac{1}{t}\int_{0}^{t}ds \hspace{2pt}\eta\left(s \hspace{1.5pt}\right)
\end{equation}
\begin{equation}\label{eta}
\langle \eta \rangle = \frac{1}{\mathcal{Z}}\int_{\mathbb{R}} \prod_{k=1}^{N} \left(dQ_{k}\hspace{1pt} dP_{k}\right)\hspace{1.5pt}\eta(\bm{Q},\bm{P}) e^{-\beta H(\bm{Q},\bm{P})},
\end{equation}
where $\mathcal{Z}$ is the canonical partition function
\begin{equation}\label{Z}
\mathcal{Z} = \int_{\mathbb{R}} \prod_{k=1}^{N} \left(dQ_{k}\hspace{1pt} dP_{k}\right)\hspace{1.5pt} e^{-\beta H(\bm{Q},\bm{P})}.
\end{equation}
When $\bar{\eta}(t)=\langle \eta \rangle$, the lattice is ergodic and is in equilibrium. Equation \ref{eta} has been approximated by neglecting the nonlinear term of the Hamiltonian, and as shown in the supplementary information of \cite{fpu dynamics} $\eta$ is given by
\begin{equation}\label{equip}
\langle \eta \rangle = \frac{1-\gamma}{S_{max} - S(0) }
\end{equation}
where $\gamma\approx 0.5772$ is the Euler-Mascheroni constant.

\begin{figure}[b]
	\includegraphics[width=.47\textwidth]{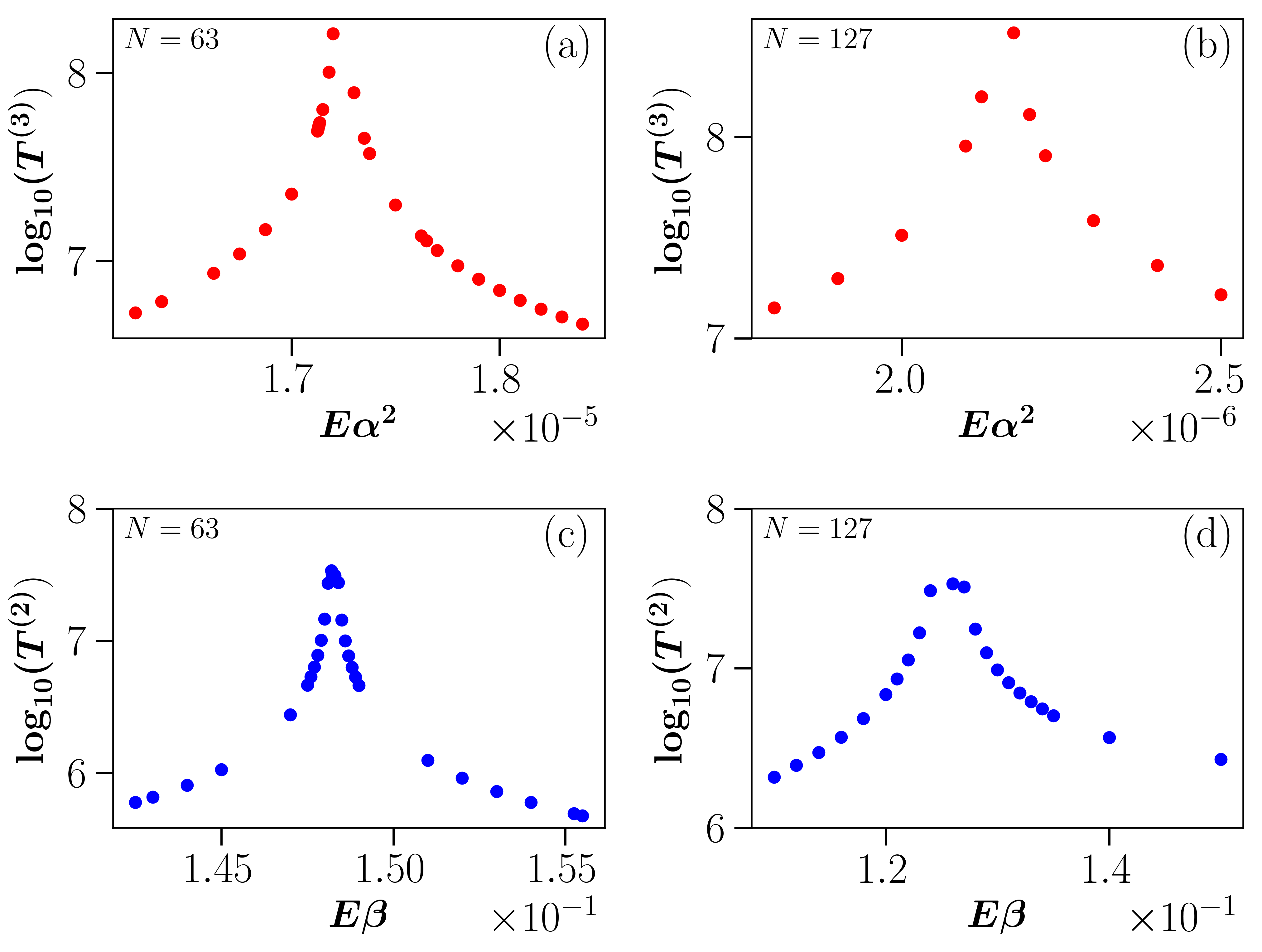}
	
	\caption{Apparent singularities in the $n$-period, which were studied in section \ref{HoR}, are shown to exist in the $\alpha$-FPUT lattice for (a) $N=63$ and (b) $N=127$ and for the $\beta$-FPUT lattice for (c) $N=63$ and (d) $N=127$. This suggests that they are not unique to a particular lattice size, and should be seen for all finite $N$. } 
	
	\label{N_sing}
\end{figure}

Fig.  \ref{SR_meta} shows $\bar{\eta}$ as a function of time for the parameters considered in Figs.  \ref{aSRbreakdown} and \ref{bSRbreakdown}. The dashed horizontal line corresponds to the ensemble average of the entropy defined by equation \ref{equip}, and $\bar{\eta}$ was obtained by numerically integrating equation \ref{eta} at each discrete time.

Fig. \ref{SR_meta}a shows the $\alpha$-FPUT lattice. Comparison between this and Fig.\ref{aSRbreakdown} shows that the breakdown of 2oRs in the $\alpha$-FPUT lattice does not correspond to equipartition. Instead, the timescale for the breakdown of the metastable state is much greater than the timescale of the formation of the mRs discussed above. In Fig, \ref{SR_meta}a, the plateau on $\bar{\eta}(t)$ is the metastable state caused by the q-breather, which occurs at noticeably different values of $\bar{\eta}$ due to the difference in energy between each curve. For the $\beta$-FPUT lattice, however, comparison of Fig. \ref{SR_meta}b and Fig. \ref{bSRbreakdown} shows that the breakdown for 2oRs corresponds to a breakdown of the metastable state, causing the lattice to approach equilibrium. The metastable plateau on $\bar{\eta}(t)$ occurs at nearly the same value of $\bar{\eta}(t)$, which is a reflection of the $2oR$ breakdown time's sensitivity on $E\beta$. The small time step size required in Fig. \ref{bSRbreakdown}, to produce Fig. \ref{SR_meta}b means that achieving $\bar{\eta}(t)=\langle \eta \rangle$ is outside our computational resources. Despite this, we feel confident that the system will achieve equipartition given enough time, as, in many previous studies (see, for example \cite{Dan camp flach}) the breakdown of the metastable state has always corresponded to a monotonically decreasing $\bar{\eta}(t)$ until $\bar{\eta}(t)=\langle \eta \rangle$. 
The lack of an abrupt breakdown of the SR in the $\alpha$-FPUT lattice may suggest some sort of intermediate phase arising from the transition to strong chaos that is not well understood. It is well known that the 3-particle $\alpha$-FPUT lattice with periodic boundary conditions can be transformed into the celebrated Hénon–Heiles system \cite{ford,Toda_HH}, which exhibits a mixed phase space where there exist chaotic and regular regions \cite{3Beta Note}. Nonetheless, based on recent results \cite{Danieli2018}, we expect the $\alpha$ model also to go eventually to equilibrium.

\section{Dependence on Lattice Size}\label{N}
Thus far all our results have been for a lattice size  $N=31$. In this section, we will show that our results are valid for larger lattice sizes by obtaining similar results for $N=63$ and $N=127$. Considering equation \ref{HS_T} and taking the linear approximation to the nonlinear frequencies while expanding the linear frequencies for large $N$ shows that $T \propto (N+1)^{3}$. Therefore, looking at larger lattices requires longer time runs to see HoRs which with the larger lattices amount to much longer computer-time runs. This makes their study very time consuming and since we find that these studies present no new results or insights. This justifies our focus on $N=31$ in most of the studies presented here.  

\begin{figure}[t!]
	\includegraphics[width=.47\textwidth]{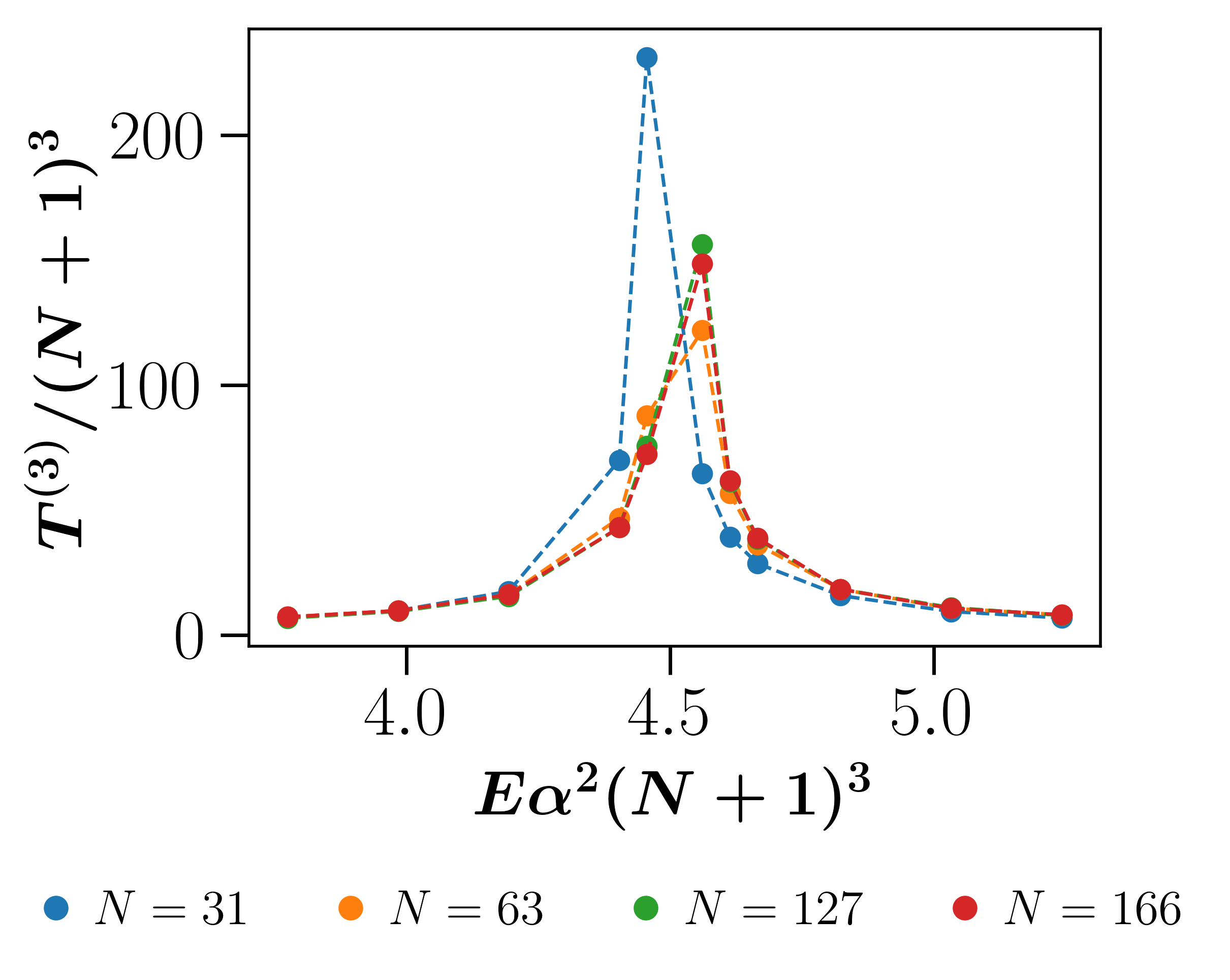}
	
	\caption{Shows a singularity in the 3-period of the $\alpha$-FPUT lattice, with $N=31$, $N=63$, $N=127$,and $N=166$. This shows the generality of the observed singularities. } 
	
	\label{rescale_sing}
\end{figure}

Figure \ref{N_sing} shows the existences of the apparent-singularities for both lattices with $N=63$ and $N=127$ in both the $\alpha$ and $\beta$-FPUT lattices. Like for $N=31$, no singularities were observed for the 2-period in the $\alpha$-FPUT lattice. 

In the $\alpha$-FPUT lattice, a natural rescaling of time is to go from $t$ to $t/(N+1)^{3}$ because $T \propto (N+1)^{3}$, as discussed above. In the KdV equation, it has been found numerically by Zabusky \cite{htf zubusky} and analytically by Toda \cite{Toda_rescale} that the rescaled FPUT-recurrence times in the $\alpha$-FPUT lattice are dependent only on the parameter $R=E\alpha^{2}(N+1)^{3}$. This parameter dependence has been previously studied by Lin et al. \cite{Lin scaling, lin}, for the FPUT-recurrence times (1oRs) in the $\alpha$-FPUT lattice, but we have found it to also work well for HoRs. These previous studies concluded that Zabusky's and Toda's parameter works well on FPUT recurrence timescales if $R^{1/3}\lesssim(N+1)$. We have found this exact inequality to not be the case for the timescales HoRs are seen. However, we did find that the idea of larger values of $R$ require larger values of $N$ for the rescaling to work well to still be true.

\begin{figure}[t!]
	\includegraphics[width=.47\textwidth]{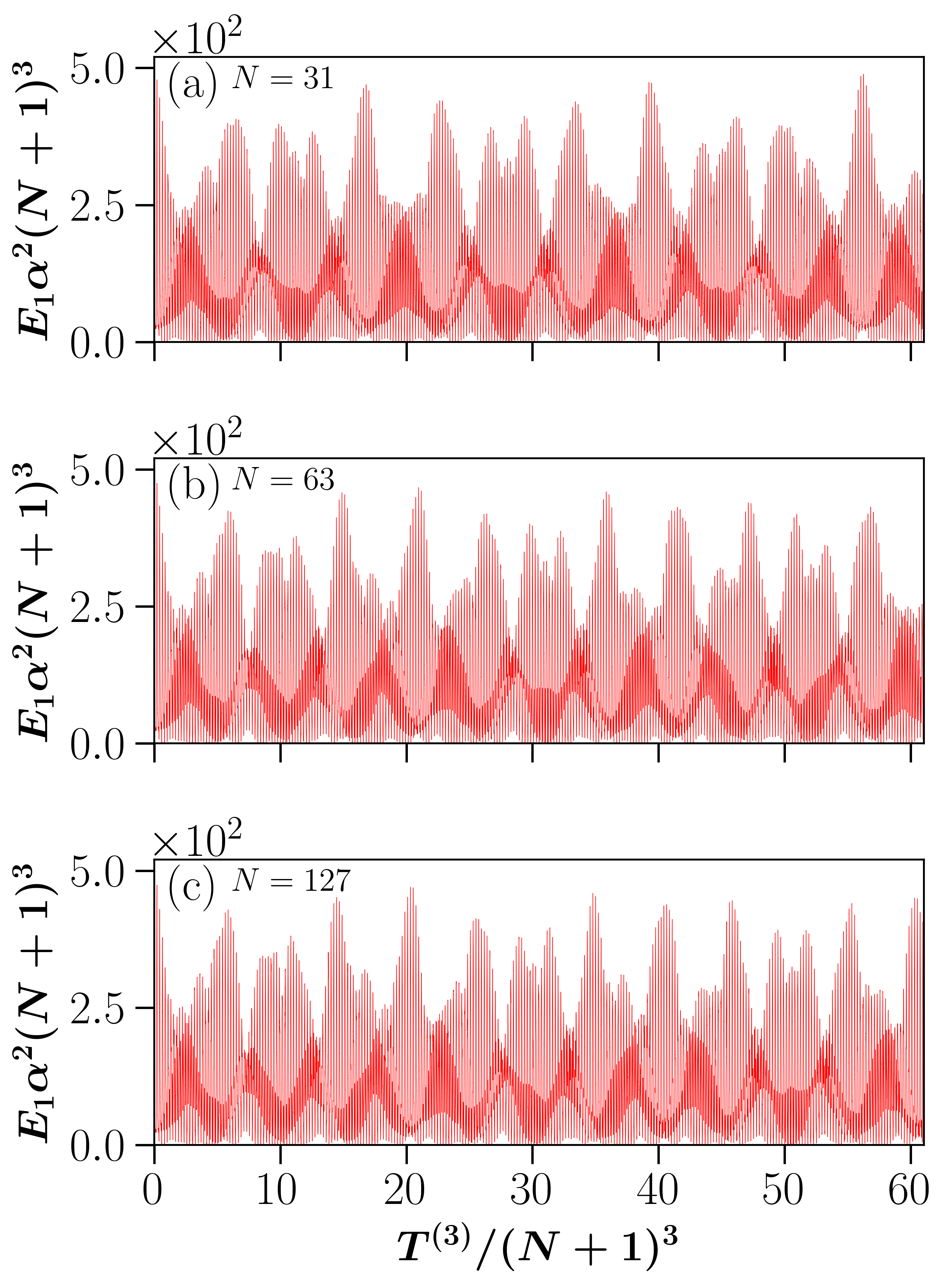}
	
	\caption{The breakdown mechanism of 2oRs in the $\alpha$-FPUT lattice, using Zabusky's and Toda's parameter $E\alpha^{2}(N+1)^{3}=491.52$ for (a) $N=31$, (b) $N=63$, and (c) $N=127$. } 
	
	\label{a_N_brk}
\end{figure}
\begin{figure}[t!]
	\includegraphics[width=.47\textwidth]{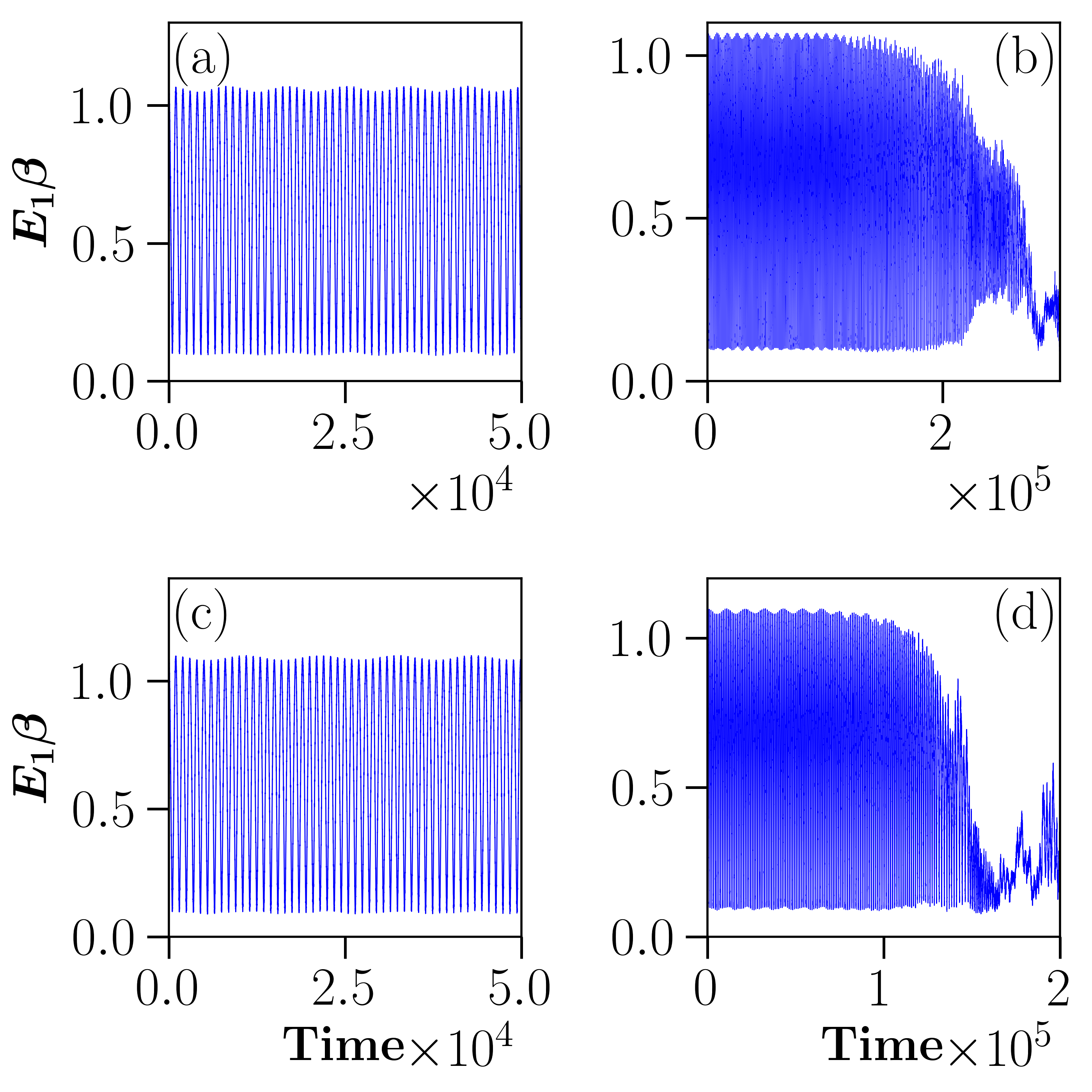}
	
	\caption{The breakdown of 2oRs in the $\beta$-FPUT lattice for $N=18$ for (a-b) $E\beta=1.07$ and (c-d) $E\beta=1.1$. As
		in our study of $N=31$, the 2oRs breakdown is associated with a breakdown of the metastable state as energy floods from the first modes into the other normal modes.} 
	
	\label{b_N_brk}
\end{figure}

Fig. \ref{rescale_sing} shows the singularity of Fig. \ref{N_sing}b with different $N$ and the parameter $E\alpha^{2}(N+1)^{3}$ fixed. Firstly, all the $N$ shown exhibit the apparent singularity which was found for $N=31$. For smaller values of $T^{(3)}/(N+1)^{3}$, $N=31$ agrees just as well as $N=166$. However, for larger values, near the peak of the singularity, small $N$ disagree slightly with larger $N$, where it seems that the center of the singularity for small $N$ is slightly shifted from the center of larger $N$. However, this disagreement for smaller lattice sizes is expected, as longer time runs will amplify any disagreements that were not noticeable for shorter times. 

In Zabusky and Kruskal's continuum limit of FPUT lattices, the $\beta$-FPUT lattice becomes the modified Korteweg-de Vries (mKdV) equation. Since some of the exact solutions of the mKdV equation are exponentially unstable \cite{mKdV} and FPUT recurrences cannot be studied simply in the continuum equation. Accordingly, there is no obvious a single parameter that describes FPUT recurrences in the $\beta$-FPUT lattice. We have looked into the parameter $E\beta(N+1)$ derived by De Luca et al. \cite{De Luca}, but it does not describe the rescaled HoRs times well for varying lattice sizes. This could be because the parameter was derived for a 4-mode subsystem of the $\beta$-FPUT Hamiltonian, and for large lattice sizes, there are many more than four active modes present. 

In terms of the breakdown mechanisms, we have found that they do not change with different lattice sizes. The $\alpha$-FPUT lattice 2oRs still breakdown by losing their shapes with increasing energy, while the $\beta$-FPUT lattice 2oRs still abruptly breakdown as energy leaves the 1st mode, diffusing into other normal modes. The 2oRs losing shape is shown for the $\alpha$-FPUT lattice for $N=63$ and $N=127$ in Fig.  \ref{a_N_brk}, with Toda's and Zabusky's parameter $R=491.52$. The values in Fig. \ref{a_N_brk}a are the same as shown by Fig.  \ref{aSRbreakdown}a. While there are subtle differences between the \ref{a_N_brk}a-c, they are most likely caused by $N$ being too small for lattices like $N=31$ at this value of $R$. Nevertheless, the general behavior of mini-recurrences destroying the shape of 2oRs is still present. For the $\beta$-FPUT lattice, the abrupt breakdown of 2oRs is shown in Fig. \ref{b_N_brk} for $N=18$. We have observed the breakdown for larger system sizes, but have not been able to reduce the time step without changing the breakdown time. Nevertheless, the confident observation of the breakdown mechanism in $N=18$ suggests the universality of the 2oR breakdown in the $\beta$-FPUT lattice.

\section{Discussion and Conclusions}

While FPUT recurrences and the questions they raise about the approach to equilibrium continue to be an active area of research today, the nature and existence of super-recurrences themselves have been the focus of few studies. Using computer-based simulations, we have investigated the existence of super-recurrences and their higher-order counterparts (HoRs) in the $\alpha$ and $\beta$-FPUT lattices. These HoRs include the previously known super-recurrences but also involve others, such as “super-super-recurrences”. Our three primary results are on the scaling of the HoRs periods with $E\alpha^{2}$ and $E\beta$, the nature of the breakdown mechanisms of super-recurrences changes with increasing $E\alpha^{2}$ and $E\beta$, and how this breakdown relates to the thermalization in the system.

The HoR periods were found to scale non-trivially due to the existence of apparent singularities, which we conjecture are caused by nonlinear resonances. These apparent singularities have been observed in both lattices, but we have not observed any in the 2-period of the $\alpha$-FPUT lattice. In both lattices, we also showed that as the values $E\alpha^{2}$ and $E\beta$ approach the center of the apparent singularity, one finds a nested assortment of HoRs, with periods smaller than the $n$-period in which the singularity exists and modulations smaller than those of the HoRs described by the $n$-period.

Furthermore, we argued that when the system thermalizes there must have been a breakdown of FPUT recurrences and therefore any other HoRs beforehand. Thus, with the strong stochastic threshold in mind, we increased the values of $E\alpha^{2}$ and $E\beta$  to see the effects on the breakdown of the 2oRs. Doing this showed that the breakdown mechanisms of the 2oRs differ between the two lattices. The breakdown of 2oRs in the $\beta$-FPUT lattice occurs abruptly, and before this breakdown the shape of the 2oRs  is normal. For the $\alpha$-FPUT lattice, the 2oRs breakdown by losing their shape until they no longer form. Also, using the spectral entropy, we showed that the breakdown of 2oRs in the $\beta$-FPUT lattice is associated with the system’s approaching equilibrium, while in the $\alpha$-FPUT model, even after the 2oRs have broken down by losing their shape until they no longer form, the system is still not at equilibrium.

In our study of HoRs, we have fixed our initial state to only include the first mode. To the best of our knowledge, there have been no studies looking into other long-wavelength initial states. It might be very interesting to study the n=2 mode in the beta lattice, when then only the even modes would be excited. Therefore, it would be interesting to see if one could observe the same phenomena observed in this paper in other long-wavelength initial states, such as initial even normal mode states in the $\beta$-FPUT lattice and linear combinations of even and odd normal mode states. Also, even for fundamental initial modes, we only studied the HoRs in the 1st normal mode. Looking at Figs \ref{arecurrences}b and \ref{brecurrences}b, one can see that some of the other modal energies also exhibit super-recurrences. Zabusky first showed that the $\text{k}^{\text{th}}$ normal mode has a peak in its energy at $t \sim T_{1oR}/k$ \cite{zabusky}. It is surprising that the other normal mode HoRs do not follow this same trend as well. 

The interesting differences between the $\alpha$ and $\beta$-FPUT lattices provide further evidence of  the subtleties involved in the approach to equilibrium at low energies in classical many-body systems.

\section{Acknowledgements}
We would like to acknowledge financial support from Boston University's Undergraduate Research Opportunities Program. We would like to also thank Gino Biondini for providing helpful resources.

\appendix
\section{Information on Numerics}
\subsection{$SABA_{2}C$ integrator}
If $H(\boldsymbol{p}(t), \boldsymbol{q}(t))$ is a Hamiltonian system with N degrees of freedom, then a vector $\boldsymbol{x}(t)=\left(\boldsymbol{p}(t), \boldsymbol{q}(t)\right)=\left(p_{1}(t),...,p_{N}(t),q_{1}(t),...,q_{N}(t)\right)$ describes the system's state at time $t$, where $q_{n}(t)$ and $p_{n}(t)$ are the canonical coordinates and momenta. The time evolution of the initial state is determined by Hamilton's equations,
\begin{equation} \label{hamilton's equations}
\frac{d p_{n}}{dt}=-\frac{\partial H}{\partial q_{n}}, \hspace{10pt} \frac{d q_{n}}{dt}=\frac{\partial H}{\partial p_{n}}.
\end{equation}
Defining the Poisson bracket of two functions $f$ and $g$ to be
\begin{equation}
\{f,g\} = \sum_{n=1}^{N}\left(\frac{\partial f}{\partial p_{n}}\frac{\partial g}{\partial q_{n}}-\frac{\partial f}{\partial q_{n}}\frac{\partial g}{\partial p_{n}}\right),
\end{equation}
equation \ref{hamilton's equations} then takes the form
\begin{equation}\label{hamiltons equations poisson brackets}
\frac{d\boldsymbol{x}(t)}{dt}=\{H,\boldsymbol{x}\}.
\end{equation}
Defining the differential operator $\hat{L}_{\xi}f\equiv \{\xi,f\}$, equation \ref{hamiltons equations poisson brackets} becomes
\begin{equation} \label{hamiltons equations diff op}
\frac{d\boldsymbol{x}(t)}{dt}=\hat{L}_{H}\boldsymbol{x}(t),
\end{equation}
and thus,
\begin{equation} \label{heq sln}
\boldsymbol{x}(t)=e^{t\hat{L}_{H}}\boldsymbol{x}(0)=\sum_{n=0}^{\infty}\frac{t^{n}}{n!}\hat{L}_{H}^{n}\boldsymbol{x}(0).
\end{equation}

Laskar and Robutel \cite{Laskar and Robutel}, presented a symplectic integration scheme for solving perturbed Hamiltonians of the form $H=A+\epsilon B$, where $A$ and $B$ are both integrable. To integrate from $t$ to $t+\tau$, the authors use the Campbell-Baker-Hausdorf theorem to approximate the evolution operator in \ref{heq sln},
\begin{equation}\label{CBH}
e^{\tau \hat{L}_{H}}\approx e^{\tau \hat{L}_{A}}e^{\tau \hat{L}_{\epsilon B}}=\prod_{j=1}^{n}e^{c_{j}\tau \hat{L}_{A}}e^{d_{j}\tau \hat{L}_{\epsilon B}}
\end{equation}
where $\sum c_{j}= \sum d_{j}=1$ and the individual terms are chosen to improve the order of error for the scheme.

The authors then developed the $SABA_{2}$ integrator:
\begin{equation}
SABA_{2} = e^{c_{1} \tau \hat{L}_{A}} e^{d_{1} \tau \hat{L}_{\epsilon B}} e^{c_{2} \tau \hat{L}_{A}} e^{d_{1} \tau \hat{L}_{\epsilon B}} e^{c_{1} \tau \hat{L}_{A}},
\end{equation}
where $c_{1}=\frac{1}{2}( 1-\frac{1}{\sqrt{3}})$, $c_{2}=\frac{1}{\sqrt{3}}$, and $d_{1}=\frac{1}{2}$, which produces an error of order $\tau^{4}\epsilon+\tau^{2}\epsilon^{2}$. Notice that $c_{1}+c_{1}+c_{2}=1$ and $d_{1}+d_{1}=1$, as required by equation \ref{CBH}. It can be improved by introducing a correction term, $C=\{\{A,B\},B\}$, that eliminates the $\tau^{2}$ dependence in the error. This corrected version becomes
\begin{equation}
SABA_{2}C=e^{-\tau^{3}\epsilon^{2}\frac{g}{2} \hat{L}_{C}}(SABA_{2})e^{-\tau^{3}\epsilon^{2}\frac{g}{2} \hat{L}_{C}}
\end{equation} 
where $g=\frac{2-\sqrt{3}}{24}$. The error term after this correction is now of order $\tau^{4}\epsilon+\tau^{4}\epsilon^{2}$.

\subsection{Integrating the FPUT lattice}
A more general form of equations \ref{afpuham} and \ref{bfpuham} is given by 
\begin{equation} \label{fpu hamiltonian}
H_{\chi}(\boldsymbol{q},\boldsymbol{p}) = \sum_{n=1}^{N}\frac{p_{n}^{2}}{2} + \sum_{n=0}^{N}\frac{1}{2}\Delta q_{n}^{2}+\frac{\chi}{u}\Delta q_{n}^{u},
\end{equation}
where $\Delta q_{n}\equiv q_{n+1}-q_{n}$. Equation \ref{fpu hamiltonian} can be separated into the form $H=A+\epsilon B$ by,
\begin{equation}
A =\sum_{n=1}^{N}\frac{p_{n}^{2}}{2}, \hspace{3pt}
B =\sum_{n=0}^{N}\frac{1}{2}\Delta q_{n}^{2}+\frac{\chi}{u}\Delta q_{n}^{u}, \hspace{3pt}
\epsilon =1. \\
\end{equation}
We find the correction term $C$, to be 
\begin{align}
\begin{split}
C&=\{\{A,B\},B\}=\sum_{n=1}^{N}\frac{\partial B}{\partial q_{n}} \frac{\partial B}{\partial q_{n}} \\
&=\sum_{n=1}^{N}(\Delta q_{n-1}-\Delta q_{n} +\chi(\Delta q_{n-1}^{u-1}-\Delta q_{n}^{u-1}))^{2}
\end{split}
\end{align}

To demonstrate the explicit form the operators  $e^{\tau \hat{L}_{A}}$,$e^{\tau \hat{L}_{B}}$, and $e^{\tau \hat{L}_{C}}$ take in the FPUT system, let us take the set $(q_{n}(t),p_{n}(t))\equiv (q_{n},p_{n})$ and  $(q_{n}(t+\tau),p_{n}(t+\tau))\equiv(\tilde{q}_{n},\tilde{p}_{n})$. For $X\in\{A,B,C\}$,
\begin{equation}
\hat{L}_{X}q_{n}= \frac{\partial X}{\partial p_{n}},\hspace{5pt} \hat{L}_{X}p_{n}= -\frac{\partial X}{\partial q_{n}}
\end{equation}
and one can easily see that $\hat{L}_{X}^{j}q_{n}=\hat{L}_{X}^{j}p_{n}=0$ for $j\ge 2$, and therefore the operator $e^{\tau \hat{L}_{X}}\equiv 1+\tau \hat{L}_{x}$. The results of the operators acting on the canonical coordinates and momenta are given bellow.

\begin{equation}
\begin{tabular}{c}
{\ul $e^{\tau \hat{L}_{A}}$} \\
$\tilde{q}_{n}=q_{n}+\tau p_{n}$ \\
$\tilde{p}_{n}=p_{n}$ \\                 
\end{tabular}
\end{equation}

\begin{equation}
\begin{tabular}{c}
{\ul $e^{\tau \hat{L}_{B}}$} \\
$\tilde{q}_{n}=q_{n}$ \\
$\tilde{p}_{n} =p_{n}+\tau(\Delta q_{n}-\Delta q_{n-1}+\chi(\Delta q_{n}^{u-1}-\Delta q_{n-1}^{u-1}))$ \\                 
\end{tabular}
\end{equation}

\begin{equation}
\begin{tabular}{c}
{\ul $e^{\tau \hat{L}_{C}}$} \\
$\tilde{q}_{n}=q_{n}$ \\
$\tilde{p}_{1}= p_{1}+2\tau \left(\Delta q_{1}-\Delta q_{2}+\chi(\Delta q_{1}^{u-1}-\Delta q_{2}^{u-1})\right) $\\
$\times \left(1+\chi(u-1)\Delta q_{1}^{u-2}\right) $\\
$+ 2\tau \left(q_{2}-2q_{1}+\chi(\Delta q_{1}^{u-1}-q_{1}^{u-1})\right) $\\
$\times\left(2+\chi(u-1)\left((q_{1})^{u-2}+\Delta q_{1}^{u-2}\right)\right) $\\

$\tilde{p}_{n=2,...,N-1}=p_{n}+2\tau(\Delta q_{n}-\Delta q_{n+1}+\chi(\Delta q_{n}^{u-1}\hspace{-4pt}-\hspace{-2pt}\Delta q_{n+1}^{u-1})) $\\
$\times \left(1+\chi(u-1)\Delta q_{n}^{u-2}\right) $\\
$+2\tau\left(\Delta q_{n}-\Delta q_{n-1}+\chi(\Delta q_{n}^{u-1}-\Delta q_{n-1}^{u-1})\right) $\\
$\times\left(2+\chi(u-1)\left(\Delta q_{n-1}^{u-2}+\Delta q_{n}^{u-2}\right)\right) $\\
$+2\tau\left(\Delta q_{n-2}-\Delta q_{n-1}+\chi(\Delta q_{n-2}^{u-1}-\Delta q_{n-1}^{u-1})\right) $\\
$\times\left(1+\chi(u-1)\Delta q_{n-1}^{u-2}\right) $\\
$\tilde{p}_{N}= p_{N}+ 2\tau \left(q_{N-1}-2q_{N}+\chi((-q_{N})^{u-1}-\Delta q_{N-1}^{u-1})\right) $\\
$\times\left(2+\chi(u-1)\left(\Delta q_{N-1}^{u-2}+(-q_{N})^{u-2}\right)\right)$ \\
$+2\tau \left(\Delta q_{N-2}-\Delta q_{N-1}+\chi(\Delta q_{N-2}^{u-1}-\Delta q_{N-1}^{u-1})\right)$ \\
$\times\left(1+\chi(u-1)\Delta q_{N-1}^{u-2}\right)$\\
\end{tabular}
\end{equation}

\subsection{Relative Energy Error}
\begin{figure}[b]
	\includegraphics[width=.47\textwidth]{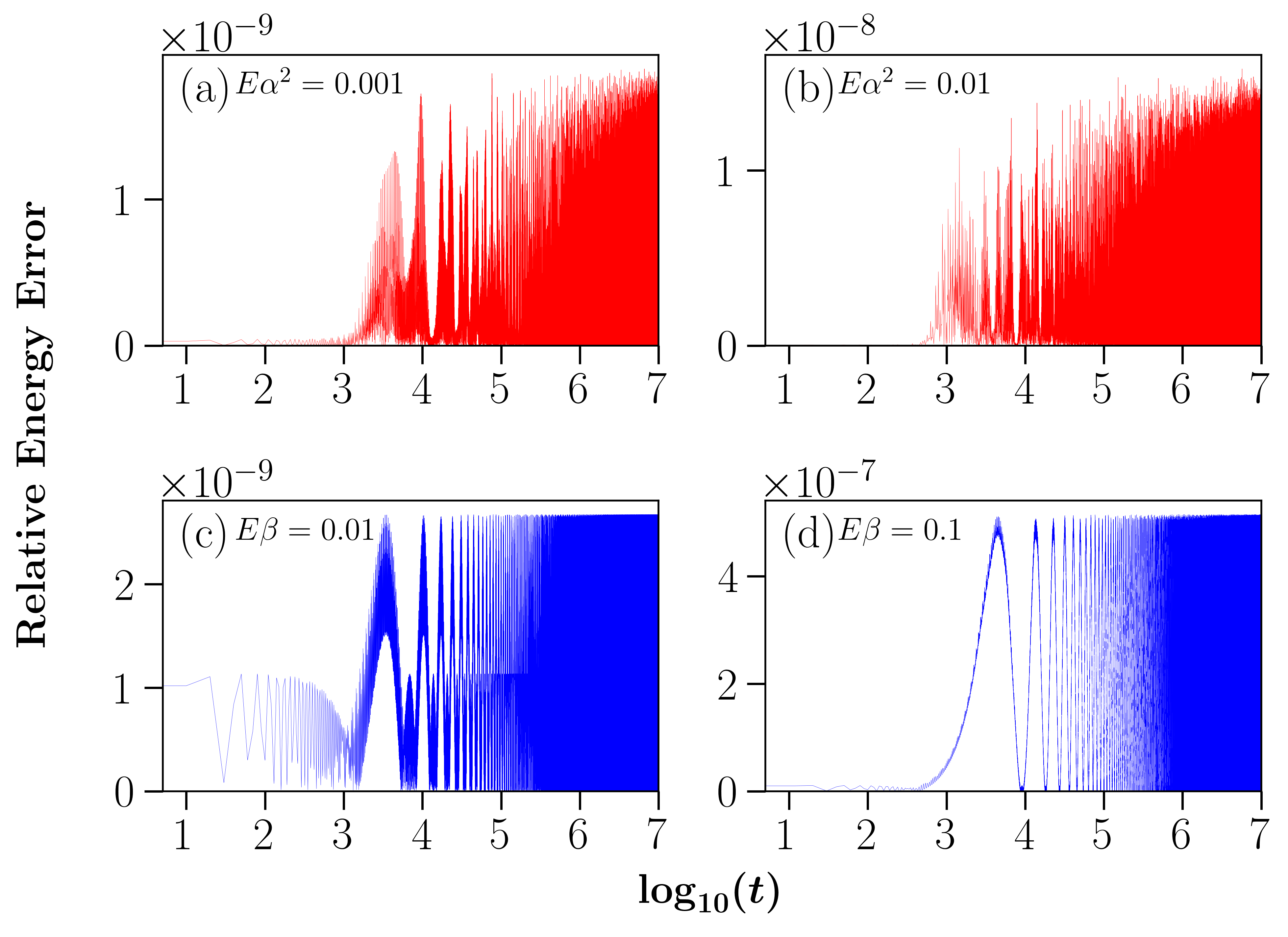}
	
	\caption{Shows the relative energy error $\mid E(t)/E(0)-1 \mid$ for the $\alpha$-FPUT lattice with (a) $E\alpha^{2}=0.001$ and (b) $E\alpha^{2}=0.01$, and also for the $\beta$-FPUT lattice at (c) $E\beta=0.01$ and (d) $E\beta=0.1$.} 
	
	\label{energy_error}
\end{figure}

We show how well the integrator conserves energy by plotting the relative energy error as a function of time, $\Delta E=\mid E(t)/E(0)-1 \mid$ for different values of $E\alpha^{2}$ and $E\beta$ with $N=31$ and $\tau=0.1$. Fig. \ref{energy_error} shows that the integrator produces a relative energy error dependent on the value of $E\alpha^{2}$ and $E\beta$. Fig. \ref{energy_error}a and \ref{energy_error}b show the $\alpha$-FPUT lattice at (a) $E\alpha^{2}=0.001$ and (b) $E\alpha^{2}=0.01$, whereas Fig. \ref{energy_error}c and \ref{energy_error}d show the $\beta$-FPUT lattice at (a) $E\beta=0.01$ and (b) $E\beta=0.1$. Both lattices show that the integrator conserves energy less well at larger values of $E\alpha^{2}$ and $E\beta$. This can be understood due to the Lyapunov exponent of both lattices increasing as a function of energy \cite{pettini}.

\subsection{Time Reversal Tests}
The FPUT equations are classical Newtonian equations of motion and thus are invariant under time reversal ($t \rightarrow -t$). Therefore, if the going forward in time from $t=0$ to say $t=\mathcal{T}$ and then stopping, reversing time and then going backward in time from $t=\mathcal{T}$ back to $t=0$, all the energy should return to the initial state. However, computers have a finite capacity to store information and integrators are not perfect, so the energy is never perfectly returned on a computer. Nevertheless, a way to verify the correctness of numerical results is to preform these time reversals and see the relative error of energy returned to the initial state. The results of this, for the calculations shown in some of our figures, are shown in table \ref{tr}. $E^{\rightarrow}$ denotes the initial energy when the forward dynamics and $E^{\leftarrow}$ is the energy returned to the initial state in the backwards dynamics.

\begin{table}[H]\label{tr}
	\begin{tabular}{cccc}
		{\ul \textbf{Figure}} & {\ul \textbf{$\tau$}} & {\ul \textbf{Precision}} & {\ul \textbf{$E^{\rightarrow}/E^{\leftarrow}-1$}} \\
		\ref{tmbrecurrences}a \& \ref{tmbrecurrences}c  & 0.001          & Double                   & $1.28786\times 10^{-14}$                          \\
		
		\ref{aHoR_existence}                    & 0.1                   & Double                   & $5.79536\times 10^{-14}$                          \\
		\ref{bHoR_existence}                    & 0.1                   & Double                   & $7.93143\times 10^{-13}$                          \\    
		\ref{aHoR_nested_HoR}a-b                     & 0.1                   & Double                   & $-4.20108\times 10^{-13}$                          \\     
		\ref{aHoR_nested_HoR}c-d                     & 0.1                   & Double                   & $-1.6228\times 10^{-12}$                          \\     
		\ref{bHoR_nested_HoR}a-b                     & 0.1                   & Double                   & $6.67688\times 10^{-13}$                          \\     
		\ref{bHoR_nested_HoR}c-d                     & 0.1                   & Double                   & $4.85234\times 10^{-12}$                          \\         
		\ref{aSRbreakdown}b \& \ref{aSRzoom}                        & 0.1                   & Double                 & $2.6823\times 10^{-13}$                          \\      
		\ref{aSRbreakdown}c                     & 0.001                   & Double                   & $-7.66814\times 10^{-8}$                          \\                     
	\end{tabular}
	\caption{Time Reversal Test Results}
\end{table}


\begin{thebibliography}{9}
\bibitem{note on N}
	Older literature has used N-1 active particles which can be confusing as N no longer refers to the number of degrees of freedom. In this paper, and in most modern day literature, N refers to the number of active particles, which is usually restricted to a power of 2 minus one or a prime number minus one to avoid internal resonance between normal modes \cite{Hemmer, Ford internal res}.
\bibitem{3Beta Note}
	Amusingly, the 3-particle $\beta$-FPUT with periodic boundary conditions turns out to actually be an integrable system \cite{Feng}.
\end{thebibliography}
\end{document}